\documentclass[pre,amsmath,preprint]{revtex4}
\usepackage{bm}
\usepackage{amssymb}
\usepackage{graphicx}
\usepackage{subfigure}
\DeclareMathOperator{\sinc}{sinc}
\date{\today}
\begin{document}

\title{Non-monotonic displacement distribution of active random walkers}
\author{Eial Teomy}
 \email{eialteom@gmail.com}
\affiliation{Institute for Physics \& Astronomy, University of Potsdam, Karl-Liebknecht-Stra{\ss}e 24/25, D-14476 Potsdam-Golm, Germany}
\author{Yael Roichman}
 \email{roichman@tau.ac.il}
 \affiliation{School of Chemistry and School of Physics and Astronomy, Tel Aviv University, Tel Aviv 69978, Israel}
 \author{Yair Shokef}
 \email{shokef@tau.ac.il}
 \affiliation{School of Mechanical Engineering and Sackler Center for Computational Molecular and Materials Science, Tel Aviv University, Tel Aviv 69978, Israel}

\begin{abstract}
We consider a simple model for active random walk with general temporal correlations, and investigate the shape of the probability distribution function of the displacement during a short time interval. We find that under certain conditions the distribution is non-monotonic and we show analytically and numerically that the existence of the non-monotonicity is governed by the walker's tendency to move forward, while the correlations between the timing of its active motion control the magnitude and shape of the non-monotonicity. In particular, we find that in a homogeneous system such non-monotonicity can occur only if the persistence is strong enough.
\end{abstract}

\maketitle

\section{Introduction}

The active and passive motion of biological cells and of their components is a complicated out of equilibrium process, which occurs due to many factors, most of them still far from being well understood \cite{Groot2005,Hofling2014,Bechinger2016,Metzler2016,Hakim2017,Norregaard2017}. This motion has been investigated both at the single-body level \cite{Hasnain2015,Detcheverry2015,Rupprecht2016,Detcheverry2017,Sevilla2016,Ariel2017,Fedotov2017} and at the many-body level \cite{Wioland2016,Stenhammer2017,Berthier2013,Viscek1995,Sepulveda2013,Grossmann2016,Liebchen2017,Zimmermann2016,Farhadifar2007,Staple2010,Sandor2017,Reichhardt2014,Zacherson2017,Reichhardt2014b,Lam2015,Graf2017,Illien2015}. The motions of individual cells or bacteria are modeled in various ways, which include correlations between the motions of a walker at different times. One of the most common models, motivated by experimental observations \cite{Berg1990}, is run and tumble motion \cite{Detcheverry2015,Rupprecht2016,Detcheverry2017,Reichhardt2014}, in which the walker moves in a straight line for some time, and then abruptly changes its direction. A twitching motion \cite{Zacherson2017} or motion with a self aligning director \cite{Reichhardt2014b} is captured by a one-step memory term, i.e. the velocity at each step depends on the velocity in the previous step but not on longer term memory. Correlated random walks are also used in other biological processes \cite{Ghosh2015}, such as DNA motors \cite{Schulz2011} and eye movements \cite{Hermann2017}, as well as in various fields such as polymer chains \cite{Tchen1952}, animal movement \cite{Kareiva1983}, scattering in disordered media \cite{Boguna1998}, artificial microswimmers \cite{Romanczuk2012,Ghosh2015b}, and motion in ordered media \cite{Tahir1983,Tahir1983b}. The origin of the active motion does not have to be the walker itself. It may also come from structural changes in the surrounding medium \cite{Lettinga2007,Pouget2011,Naderi2013,SonnSegev2017}.

In simple Markovian random walks there are no correlations between the motions of the walker at different times. The motion of active walkers, on the other hand, can be correlated in either space or time (or both). \textit{Spatial correlations}, i.e. the correlations between the direction of motion at subsequent points in time, are modeled by persistent motion, in which the walker prefers to continue moving in the same direction as before (or in the opposite direction in the case of anti-persistence). This preference could be either discrete, as in the persistent random walk model \cite{Goldstein1951}, or continuous, as in the Orenstein-Uhlenbeck process \cite{Uhlenbeck1930}. \textit{Temporal correlations}, i.e. correlations between the time intervals between direction changes, are modeled by non-Poissonian distributed running times, such as in the continuous time random walk model \cite{Montroll1965} or in Levy walks \cite{Shlesinger1987} and Levy flights \cite{Levy1939}.

In simple random walks, the probability distribution function (PDF) of the displacements, or the van-Hove distribution function, is Gaussian \cite{Rayleigh1880}. In many models of active matter, although the PDF might not be Gaussian, it is observed to still be monotonically decreasing with the magnitude of the displacement \cite{Nigris2017,Coding2008,Metzler2014,Sposini2018,Jakub2018,Gnacik2018}. However, there are cases in which the PDF is not monotonic and exhibits peaks, such as an active particle inside a harmonic potential well \cite{BenIsaac2015,Razin2019}, or an aging Levy walk \cite{Magdziarz2017}. In the overdamped limit, this is related to peaks that may appear in the velocity distribution due to biological activity \cite{BenIsaac2011}. We are not aware of any experimental results or other theoretical works which found such non-monotonicity.

In this paper we consider a simple model for active random walk and show that in a homogeneous system, such non-monotonic behaviour is not affected at all by temporal correlations, and it exists only if spatial correlations are positive and strong enough. Temporal correlations do affect the magnitude and the shape of the non-monotonicity, however. We also derive in this paper an explicit expression for the displacement PDF for general temporal correlations and a simple realization of the spatial correlations. 

Our results may be used to understand the microscopic processes underlying the overall observed random walk. At long times many of the details of the random motion are averaged out, but they are present in the short time behaviour. Using our results, and expanding the simple model we present here, one may look for non-monotonicity; or, if such non-monotonicity is not observed at short times, this implies a bound on the orientational correlations. By looking for peaks in the displacement distribution in experiments, one may obtain information about the directional correlations of the particles. Moreover, the heights and the separation of the peaks give further insight into the underlying microscopic processes.

The remainder of the paper is organized as follows. In Section \ref{sec_model} we introduce our model, and in Section \ref{sec_derpdf} we solve for the displacement PDF in it. Section \ref{sec_proof} contains our proof that the non-monotonicity may appear only if the persistence is positive, and Section \ref{sec_summary} summarizes the paper.

\section{Model}
\label{sec_model}

We consider a particle moving in $d$ dimensions and are interested in the probability density $P_{r}\left(\vec{r},\tau\right)$ that during a time interval $\tau$ its displacement was $\vec{r}$. We model the movement of the particle as a sum of two independent processes, a \textit{discrete} process representing the active or biological stochastic motion, and a \textit{continuous} process representing the contact of the particle with its thermal environment. Assuming that: 1) the thermal process is isotropic, 2) the directional correlations in the active motion are always relative to the current direction, and 3) the initial condition is isotropic, then when averaging over multiple particles their motion is isotropic, i.e. $P_{r}=P_{r}\left(r,\tau\right)$ depends only on the magnitude $r=|\vec{r}|$ of the displacement. 

The probability $P_{r}\left(\vec{r},\tau\right)$ is given by the following convolution of these two sources of fluctuation
\begin{align}
P_{r}\left(\vec{r},\tau\right)=\int P_{d}\left(\vec{r}',\tau\right)P_{c}\left(\vec{r}-\vec{r}',\tau\right)d\vec{r}' ,\label{pr_def}
\end{align}
where $P_{d}\left(\vec{r},\tau\right)$ is the probability that the displacement of the particle due to the discrete process after time $\tau$ is $\vec{r}$, and $P_{c}\left(\vec{r},\tau\right)$ is analogously defined for the continuous process. For simplicity, we assume that the continuous process is Gaussian, 
\begin{align}
P_{c}\left(\vec{r},\tau\right)=K_{d}\exp\left(-\frac{r^{2}}{2\alpha^{2}(\tau)}\right) ,\label{pc_def}
\end{align}
with the normalization factor $K_{d}$ in $d$ dimensions given by
\begin{align}
K_{d}=\frac{1}{2^{d/2-1}\alpha^{d}\Gamma\left(\frac{d}{2}\right)\Omega_{d}} ,\label{k_def}
\end{align}
where $\Omega_{d}$ is the surface area of a $d$-dimensional hypersphere, and $\alpha^2(\tau) = \langle r^2(\tau) \rangle$ is the mean squared displacement of the particle during a time interval $\tau$, solely due to the continuous process. Note that the dependence of $\alpha^{2}$ on the time interval $\tau$ could be non-trivial, and not necessarily diffusive. We describe here our model in terms of general spatial dimensionality $d$, but later on we will concentrate on the case $d=3$.

For the discrete process, the time intervals between hops, the distance of each hop and their direction could have any distribution, and can all be correlated in some fashion. In general, the probability $P_{d}\left(\vec{r},\tau\right)$ is given by
\begin{align}
P_{d}\left(\vec{r},\tau\right)=\sum^{\infty}_{n=0}P_{n}\left(\vec{r},\tau\right) ,\label{eq_pr1}
\end{align}
where $P_{n}\left(\vec{r},\tau\right)$ is the probability density that the particle performed $n$ hops during the time interval $\tau$ and that it moved a total distance $\vec{r}$ due to these $n$ hops. Assuming that the time intervals between hops is independent of the details of the hops (i.e. their length and direction), Eq. (\ref{eq_pr1}) may be written as
\begin{align}
P_{d}\left(\vec{r},\tau\right)=\sum^{\infty}_{n=0}q_{n}\left(\tau\right)p_{n}\left(\vec{r}\right) ,\label{pdsum}
\end{align}
where $q_{n}\left(\tau\right)$ is the probability that the particle performed $n$ hops until time $\tau$, and $p_{n}\left(\vec{r}\right)$ is the probability that the particle moved a total distance $\vec{r}$ in these hops given that it performed $n$ hops. Obviously, since the dynamics are isotropic $P_{r}$ and $p_{n}$ depend only on the magnitude $r=\left|\vec{r}\right|$ of the displacement, and not on the direction of $\vec{r}$, but we still keep the notation $\vec{r}$ to emphasize that these are probability densities with respect to the d-dimensional vector $\vec{r}$ and not with respect to the scalar $r$.

\begin{figure}
\includegraphics[width=0.6\columnwidth]{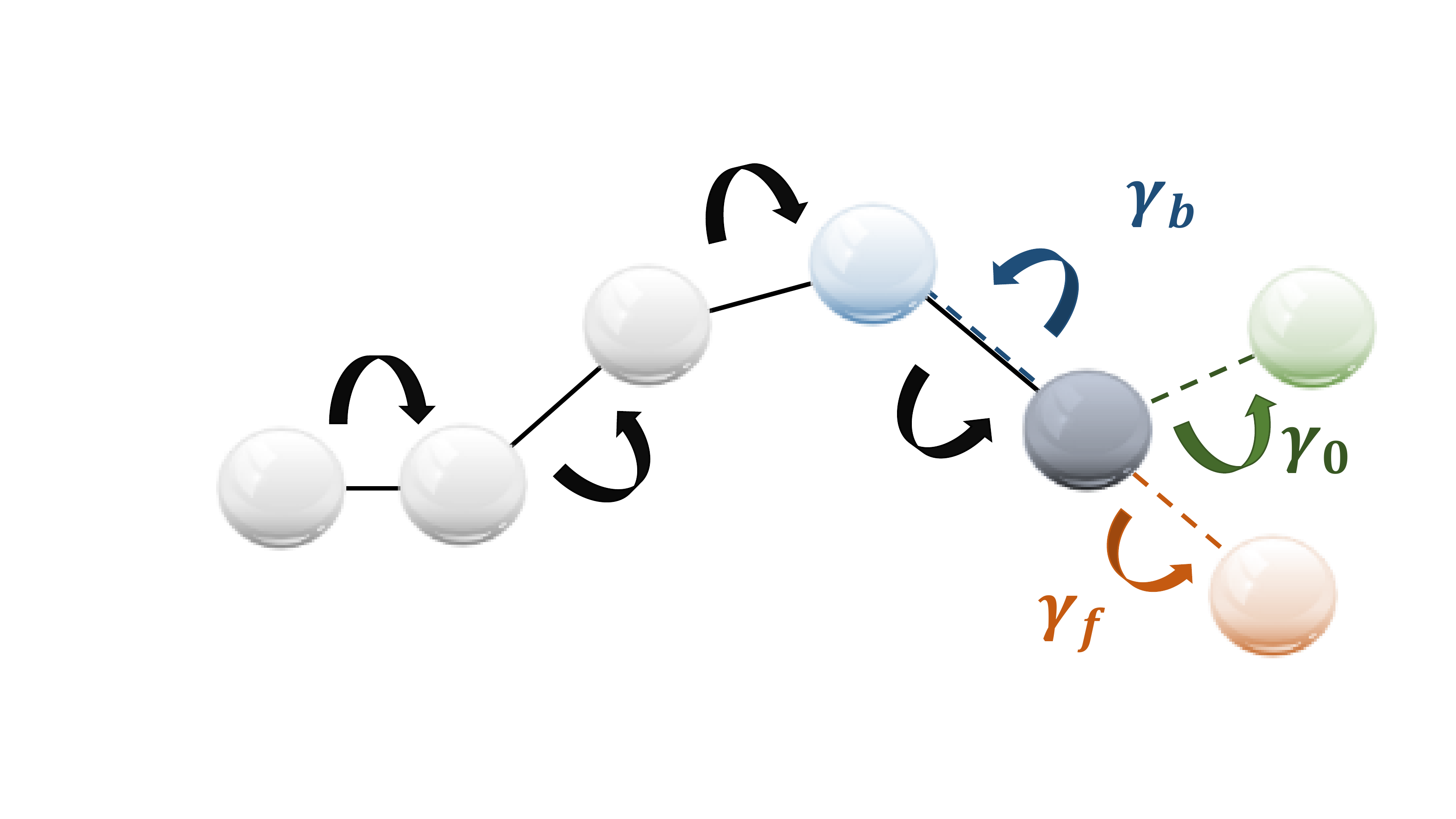}
\caption{Schematic illustration of the model showing a sequence of discrete hops. In the last step the walker reached the position marked in dark gray. In the next step, with probability $\gamma_{f}$ it will perform another step of the same magnitude and in the same direction (orange), with probability $\gamma_{b}$ it will retrace its last step (blue) and with probability $\gamma_{0}$ it will move a random distance drawn from the distribution $s\left(\ell\right)$ in a random direction (green).}
\label{schem}
\end{figure}

We will assume a simple persistent random walk model for the probability $p_{n}\left(\vec{r}\right)$, as schematically shown in Fig. \ref{schem}. At each hop, the particle performs with probability $\gamma_{0}\equiv 1-\gamma_{f}-\gamma_{b}$ an uncorrelated hop and moves in a random direction a distance $\ell$ which is drawn from some given distribution $s(\ell)$. With probability $\gamma_{f}$ the particle moves forward and repeats its previous hop (same direction and same magnitude), and with probability $\gamma_{b}$ it moves back, namely it performs the exact opposite of its previous hop (opposite direction and same magnitude).

In reality, directional correlations are not necessarily sharp as in our model. Namely, if the particle persists in its direction of motion between two subsequent hops, this need not imply that the directions of these two hops are exactly equal. We could extend our work to situations in which there is some general probability distribution function to change the direction of motion by an arbitrary angle. Similarly, the magnitudes of correlated hops are not necessarily identical, and we could allow the length of the hop to be random also if the particle hops in the same direction. However, in this paper we would like to concentrate on the simplest possible version of our model in order to obtain analytical results which remain qualitatively true even in more general cases. In particular, as we will show below, a non-monotonic behaviour of the PDF indicates positive directional correlations.

\section{Derivation of the displacement distribution}
\label{sec_derpdf}

In this section we derive the PDF of the displacements. We start by simplifying the convolution representation of the total displacement, Eq. (\ref{pr_def}), for a Gaussian continuous process and a general discrete process. Next, we consider our simple model for the discrete process, still with general step size distribution and general temporal correlations. Lastly, we choose several specific step size distributions and temporal correlations, in order to demonstrate our results.

\subsection{PDF of the total displacement for general discrete processes} 

Combining Eqs. (\ref{pr_def}-\ref{k_def}) yields
\begin{align}
P_{r}\left(\vec{r},\tau\right)=\frac{1}{2^{d/2-1}\alpha^{d}\Gamma\left(\frac{d}{2}\right)\Omega_{d}}\int P_{d}\left(\vec{r}',\tau\right)\exp\left(-\frac{\left(\vec{r}-\vec{r}'\right)^{2}}{2\alpha^{2}}\right)d\vec{r}' .
\end{align}
Assuming that $P_{d}$ is isotropic, integrating over the angles yields
\begin{align}
P_{r}\left(\vec{r},\tau\right)=\frac{1}{2^{d/2-1}\alpha^{d-2}\Gamma\left(\frac{d}{2}\right)}\int P_{d}\left(\vec{r}',\tau\right)\frac{r'^{d-2}}{r}\exp\left(-\frac{\left(r^{2}+r'^{2}\right)}{2\alpha^{2}}\right)\sinh\left(\frac{rr'}{\alpha^{2}}\right)dr' .\label{pr_2}
\end{align}
We now define the Fourier transform of $P_{d}\left(\vec{r},\tau\right)$, 
\begin{align}
&P_{d}\left(\vec{k},\tau\right) = \int d\vec{r} e^{i\vec{k}\cdot\vec{r}} \tilde{P}_{d}\left(\vec{r},\tau\right) .\label{pd_fourier}
\end{align}
Inverting the Fourier transform and integrating over the angles yields
\begin{align}
&P_{d}\left(\vec{r},\tau\right) = \frac{\Omega_{d}}{\left(2\pi\right)^{d}} \int dk k^{d-1}\sinc\left(kr\right)\tilde{P}_{d}\left(\vec{k},\tau\right) .\label{pd_fourier_2}
\end{align}
Combining Eqs. (\ref{pr_2}) and (\ref{pd_fourier_2}) yields
\begin{align}
&P_{r}\left(\vec{r},\tau\right)=\frac{\Omega_{d}}{2^{3d/2-1}\pi^{d}\alpha^{d-2}r\Gamma\left(\frac{d}{2}\right)}\times\nonumber\\
&\times\int \tilde{P}_{d}\left(\vec{k},\tau\right)k^{d-2}r'^{d-3}\sin(kr')\exp\left(-\frac{\left(r^{2}+r'^{2}\right)}{2\alpha^{2}}\right)\sinh\left(\frac{rr'}{\alpha^{2}}\right)dr'dk .\label{pr_3}
\end{align}
Integrating over $r'$ yields
\begin{align}
&P_{r}\left(\vec{r},\tau\right)=\frac{\Omega(d)\exp\left(-\frac{r^{2}}{2\alpha^{2}}\right)}{2^{d+1}\pi^{d}\left(d-2\right)ir}\times\nonumber\\
&\int^{\infty}_{0}dk P_{d}\left(\vec{k},\tau\right)k^{d-2}\left\{{}_{1}F_{1}\left[\frac{d}{2}-1,\frac{1}{2},-\frac{\left(\alpha^{2}k-ir\right)^{2}}{2\alpha^{2}}\right]-{}_{1}F_{1}\left[\frac{d}{2}-1,\frac{1}{2},-\frac{\left(\alpha^{2}k+ir\right)^{2}}{2\alpha^{2}}\right]\right\} ,\label{pr_4}
\end{align}
where ${}_{1}F_{1}$ is the confluent hypergeometric function \cite{hypergeometric}. For $d=3$, Eq. (\ref{pr_4}) reduces to
\begin{align}
P_{r}\left(\vec{r},\tau\right)=\frac{1}{2\pi^{2}r}\int^{\infty}_{0}dk P_{d}\left(\vec{k},\tau\right)k\sin\left(kr\right)\exp\left(-\frac{\alpha^{2}k^{2}}{2}\right) .\label{pr_4_d3}
\end{align}
For $d=1$ or $d=2$, Eq. (\ref{pr_4}) is not valid, since its derivation includes integration over variables that exist only in $d\geq3$. However, it is straightforward to check that for $d=1$, the equivalent result to Eq. (\ref{pr_4}) is
\begin{align}
P_{r}\left(r,\tau\right)=\frac{1}{\pi}\int^{\infty}_{0}e^{-\alpha^{2}k^{2}/2}\cos\left(kr\right)\tilde{P}_{d}\left(k,\tau\right)dk ,
\label{pr_4_d1}
\end{align}
and for $d=2$ the equivalent result is
\begin{align}
P_{r}\left(\vec{r},\tau\right)=\frac{1}{2\pi\alpha}\exp\left(-\frac{r^{2}}{2\alpha^{2}}\right)\int^{\infty}_{0}dkP_{d}\left(\vec{k},\tau\right)\int^{\infty}_{0}dxe^{-x^{2}/2}I_{0}\left(\frac{xr}{\alpha}\right)J_{0}\left(\alpha k x\right) .
\label{pr_4_d2}
\end{align}
Therefore, for any stochastic process that can be decomposed into a Gaussian continuous process and an active discrete process, by knowing the distribution of the discrete process, $P_{d}\left(\vec{r},\tau\right)$, we can perform its Fourier transform and substitute $P_{d}\left(\vec{k},\tau\right)$ in Eq. (\ref{pr_4}) to obtain $P_{r}\left(\vec{r},\tau\right)$. We will now obtain $P_{d}$ for our specific simple model for the discrete dynamics, and for it we will calculate $P_{r}$.

\subsection{The Fourier transform of the model specific discrete process}

We now consider our specific model for the discrete process, as outlined in Section \ref{sec_model}. We do not yet specify the step size distribution $s\left(\ell\right)$ and the temporal correlations encoded in $q_n(\tau)$. Separating the spatial and temporal correlations of the discrete process as in Eq. (\ref{pdsum}), we now write the probability that the particle moved a distance $\vec{r}$ given it performed $n$ hops, $p_{n}\left(\vec{r}\right)$, as an integral over all the possible lengths and directions of the last hop $n$ that the particle performed and which brought it to $\vec{r}$, 
\begin{align}
p_{n}\left(\vec{r}\right)=\int d\ell d\hat{r} p_{n,\ell\hat{r}}\left(\vec{r}\right) ,\label{pn_integral}
\end{align}
where $p_{n,\ell\hat{r}}\left(\vec{r}\right)$ is the probability density (in $\vec{r}$, $\ell$ and $\hat{r}$) that the particle is located at $\vec{r}$ after $n$ hops and that in the last hop it moved a distance $\ell$ in direction $\hat{r}$. Clearly $p_{n,\ell\hat{r}}\left(\vec{r}\right)$ includes in it the probability distribution for the direction and magnitude of the previous steps. For our model it obeys the following recursion relation
\begin{align}
p_{n+1,\ell\hat{r}}\left(\vec{r}\right) = \gamma_0 \frac{s\left(\ell\right)}{\Omega_{d}}\int d\ell' d\hat{r}' p_{n,\ell'\hat{r}'}\left(\vec{r}-\ell\hat{r}\right)  + \gamma_{f} p_{n,\ell\hat{r}}\left(\vec{r}-\ell\hat{r}\right)+ \gamma_{b} p_{n,-\ell\hat{r}}\left(\vec{r}-\ell\hat{r}\right) . \label{pnrec}
\end{align}
The first term represents the case that hop $n+1$ was chosen with a random length with probability density $s\left(\ell\right)$ and in a random direction in $d$ dimensions with a uniform azimuthal probability density $\frac{1}{\Omega_{d}}$, the second term represents the case that hop $n+1$ was identical in direction and length to hop $n$, and the third term represents the case that hop $n+1$ was the exact opposite of hop $n$. The integration variables $\ell'$ and $\hat{r}'$ in the first term are the magnitude and direction of hop $n$. 

Given Eqs. (\ref{pr_4},\ref{pr_4_d3}), to proceed with the solution we introduce the Fourier transform of $p_{n,\ell\hat{r}}\left(\vec{r}\right)$,
\begin{align}
&\tilde{p}_{n,\ell\hat{r}}\left(\vec{k}\right) = \int d\vec{r} e^{i\vec{k}\cdot\vec{r}} p_{n,\ell\hat{r}}\left(\vec{r}\right) ,\nonumber\\
&p_{n,\ell\hat{r}}\left(\vec{r}\right) = \frac{1}{\left(2\pi\right)^{d}} \int d\vec{k} e^{-i\vec{k}\cdot\vec{r}} \tilde{p}_{n,\ell\hat{r}}\left(\vec{k}\right) , \label{fourier}
\end{align}
Substituting this in the recursion relation~(\ref{pnrec}) yields
\begin{align}
\tilde{p}_{n+1,\ell\hat{r}}\left(\vec{k}\right) &= \nonumber\\
&= 
\int d\vec{r} e^{i\vec{k}\cdot\vec{r}}
\left[
\gamma_0 \frac{s\left(\ell\right)}{\Omega_{d}} \cdot \frac{1}{\left(2\pi\right)^{d}} \int d\ell' d\hat{r}' d\vec{k}' e^{-i\vec{k}'\cdot\left(\vec{r}-\ell\hat{r}\right)} \tilde{p}_{n,\ell'\hat{r}'}\left(\vec{k}'\right) \right.\nonumber\\
&\left.+\gamma_{f} \frac{1}{\left(2\pi\right)^{d}} \int d\vec{k}' e^{-i\vec{k}'\cdot\left(\vec{r}-\ell\hat{r}\right)} \tilde{p}_{n,\ell\hat{r}}\left(\vec{k}'\right) +\gamma_{b} \frac{1}{\left(2\pi\right)^{d}} \int d\vec{k}' e^{-i\vec{k}'\cdot\left(\vec{r}-\ell\hat{r}\right)} \tilde{p}_{n,-\ell\hat{r}}\left(\vec{k}'\right)\right]  \nonumber\\
&=
\gamma_0 \frac{s\left(\ell\right)}{\Omega_{d}} \int d\ell' d\hat{r}' d\vec{k}' e^{i\vec{k}'\cdot\ell\hat{r}} \tilde{p}_{n,\ell'\hat{r}'}\left(\vec{k}'\right)\delta\left(\vec{k}-\vec{k}'\right) \nonumber\\
&+ \gamma_{f} \int d\vec{k}' e^{i\vec{k}'\cdot\ell\hat{r}} \tilde{p}_{n,\ell\hat{r}}\left(\vec{k}'\right)\delta\left(\vec{k}-\vec{k}'\right) + \gamma_{b} \int d\vec{k}' e^{i\vec{k}'\cdot\ell\hat{r}} \tilde{p}_{n,-\ell\hat{r}}\left(\vec{k}'\right)\delta\left(\vec{k}-\vec{k}'\right)  \nonumber\\
&=
\gamma_0 \frac{s\left(\ell\right)}{\Omega_{d}}\int d\ell' d\hat{r}' e^{i\vec{k}\cdot\ell\hat{r}} \tilde{p}_{n,\ell'\hat{r}'}\left(\vec{k}\right)
+ \gamma_{f} e^{i\vec{k}\cdot\ell\hat{r}} \tilde{p}_{n,\ell\hat{r}}\left(\vec{k}\right)+ \gamma_{b} e^{i\vec{k}\cdot\ell\hat{r}} \tilde{p}_{n,-\ell\hat{r}}\left(\vec{k}\right) ,
\end{align}
where the first equality is substitution of (\ref{pnrec}) and (\ref{fourier}), in the second equality we integrated over $\vec{r}$, and in the third equality we integrated over $\vec{k}'$. 

This may be written in operator form as
\begin{align}
\left|\tilde{p}_{n+1}\left(\vec{k}\right)\right\rangle = \left( \gamma_{0}{\cal M}_{0}+ \gamma_{f}{\cal M}_{f} + \gamma_{b}{\cal M}_{b} \right) \left|\tilde{p}_{n}\left(\vec{k}\right)\right\rangle ,
\end{align}
where $\left|\tilde{p}_{n}\left(\vec{k}\right)\right\rangle$ is an infinite-dimensional vector whose entries are $\tilde{p}_{n,\ell\hat{r}}$, and the linear operators ${\cal M}_{0},{\cal M}_{f}$ and ${\cal M}_{b}$ are defined by the following matrix elements
\begin{align}
&\left[{\cal M}_{0}\right]_{\ell\hat{r},\ell'\hat{r}'} = \frac{s\left(\ell\right)}{\Omega_{d}} e^{i\vec{k}\cdot\ell\hat{r}} ,\nonumber\\
&\left[{\cal M}_{f}\right]_{\ell\hat{r},\ell'\hat{r}'} = e^{i\vec{k}\cdot\ell\hat{r}} \delta\left(\ell-\ell'\right) \delta\left(\hat{r}-\hat{r}'\right) ,\nonumber\\
&\left[{\cal M}_{b}\right]_{\ell\hat{r},\ell'\hat{r}'} = e^{i\vec{k}\cdot\ell\hat{r}} \delta\left(\ell-\ell'\right) \delta\left(\hat{r}+\hat{r}'\right) .
\end{align}
Note that $\left|\tilde{p}_{n}\left(\vec{k}\right)\right\rangle$ is related to $\tilde{p}_{n}\left(\vec{k}\right)$ by the Fourier transform of Eq. (\ref{pn_integral}), or equivalently in bra-ket notation
\begin{align}
\tilde{p}_{n}\left(\vec{k}\right)=\left\langle 1\right|\left.\tilde{p}_{n}\left(\vec{k}\right)\right\rangle .
\end{align}
Thus the evolution of the distribution with the number of hops may be written as
\begin{align}
\left|\tilde{p}_{n}\left(\vec{k}\right)\right\rangle = \left( \gamma_{0}{\cal M}_{0} + \gamma_{f}{\cal M}_{f}+ \gamma_{b}{\cal M}_{b} \right)^{n-1} \left|\tilde{p}_{1}\left(\vec{k}\right)\right\rangle .
\end{align}

We find the distribution $\left|\tilde{p}_{1}\left(\vec{k}\right)\right\rangle$ after the first hop by
\begin{align}
\tilde{p}_{1,\ell\hat{r}}\left(\vec{k}\right) = \int d\vec{r} e^{i\vec{k}\cdot\vec{r}} p_{1,\ell\hat{r}}\left(\vec{r}\right) = \frac{s(\ell)}{\Omega_{d}} \int d\vec{r} e^{i\vec{k}\cdot\vec{r}} \delta\left(\vec{r}-\ell\hat{r}\right) = \frac{s(\ell)}{\Omega_{d}}e^{i\vec{k}\cdot\ell\hat{r}} .
\end{align}
Integrating over all possible lengths $\ell$ and directions $\hat{r}$ of hop $n$ we obtain
\begin{align}
\tilde{p}_{1}\left(\vec{k}\right) = \int d\ell d\hat{r} \tilde{p}_{1,\ell\hat{r}}\left(\vec{k}\right) = \int d\ell d\hat{r} \frac{s(\ell)}{\Omega_{d}} e^{i\vec{k}\cdot\ell\hat{r}}  = \int d\ell s(\ell) sinc(k\ell) \equiv f(k), \label{fdef}
\end{align}
where in the last equality we introduce the function $f(k)$, which is a transform of the single step size distribution $s(\ell)$. We note that this function is symmetric, $f(k)=f(-k)$.
Using the relations
\begin{align}
&{\cal M}_{0}\left|\tilde{p}_{1}\left(m\vec{k}\right)\right\rangle=f(mk)\left|\tilde{p}_{1}\left(\vec{k}\right)\right\rangle ,\nonumber\\
&{\cal M}_{f}\left|\tilde{p}_{1}\left(m\vec{k}\right)\right\rangle=\left|\tilde{p}_{1}\left[(1+m)\vec{k}\right]\right\rangle ,\nonumber\\
&{\cal M}_{b}\left|\tilde{p}_{1}\left(m\vec{k}\right)\right\rangle=\left|\tilde{p}_{1}\left[(1-m)\vec{k}\right]\right\rangle ,\nonumber\\
&\left\langle 1|\tilde{p}_{1}\left(m\vec{k}\right)\right\rangle=f(mk) ,\label{mb_reduce}
\end{align}
it is a matter of straightforward algebra to find the value of 
\begin{align}
\tilde{p}_{n}\left(\vec{k}\right) &= \int d\ell d\hat{r} \tilde{p}_{n,\ell\hat{r}}\left(\vec{k}\right) =
\left\langle 1\right|\left(\gamma_{0}{\cal M}_{0}+\gamma_{f}{\cal M}_{f}+\gamma_{b}{\cal M}_{b}\right)^{n-1}\left|\tilde{p}_{1}\left(\vec{k}\right)\right\rangle ,\label{pngen}
\end{align}
for any finite $n$. The evaluation can be simplified by noting that the operators satisfy
\begin{align}
&{\cal M}^{2}_{b}={\bf 1} ,\nonumber\\
&{\cal M}_{f}{\cal M}_{b}{\cal M}_{f}={\cal M}_{b} ,\nonumber\\
&{\cal M}_{0}{\cal M}_{b}{\cal M}_{0}={\cal M}_{0} ,\nonumber\\
&{\cal M}_{0}{\cal M}_{b}{\cal M}_{f}={\cal M}_{0} ,\nonumber\\
&{\cal M}_{f}{\cal M}_{b}{\cal M}_{0}={\cal M}_{0} ,
\end{align}
where ${\bf 1}$ is the identity operator. For small values of $n$, Eq. (\ref{pngen}) yields
\begin{align}
&\tilde{p}_{0}\left(\vec{k}\right)=1 ,\nonumber\\
&\tilde{p}_{1}\left(\vec{k}\right)=f_{1} ,\nonumber\\
&\tilde{p}_{2}\left(\vec{k}\right)=\gamma_{0}f^{2}_{1}+\gamma_{f} f_{2}+\gamma_{b} ,\nonumber\\
&\tilde{p}_{3}\left(\vec{k}\right)=\gamma^{2}_{0}f^{3}_{1}+2\gamma_{0}\gamma_{f}f_{1}f_{2}+\gamma_{f}^{2}f_{3}+\gamma_{b}\left(2-\gamma_{b}\right)f_{1} ,\nonumber\\
&\tilde{p}_{4}\left(\vec{k}\right)=\gamma^{3}_{0}f^{4}_{1}+3\gamma^{2}_{0}\gamma_{f}f^{2}_{1}f_{2}+\gamma_{0}\gamma^{2}_{f}\left(2f_{1}f_{3}+f^{2}_{2}\right)+\gamma^{3}_{f}f_{4}+\nonumber\\
&+\gamma_{b}\gamma_{0}\left(3+\gamma_{f}-\gamma_{b}\right)f^{2}_{1}+2\gamma_{f}\gamma_{b}f_{2}+\gamma_{b}\left(\gamma_{b}+\gamma^{2}_{f}\right) ,\nonumber\\
&\tilde{p}_{5}\left(\vec{k}\right)=\gamma^{4}_{0}f^{5}_{1}+4\gamma^{3}_{0}\gamma_{f}f^{3}_{1}f_{2}+3\gamma^{2}_{0}\gamma^{2}_{f}f_{1}\left(f_{1}f_{3}+f^{2}_{2}\right)+\nonumber\\
&+2\gamma_{0}\gamma^{3}_{f}\left(f_{1}f_{4}+f_{2}f_{3}\right)+\gamma^{4}_{f}f_{5}+\gamma_{b}\left[\gamma_{b}+2\left(1-\gamma_{b}\right)\left(\gamma_{b}+\gamma^{2}_{f}\right)\right]f_{1}+\nonumber\\
&+\gamma_{b}\gamma^{2}_{0}\left(4+2\gamma_{f}-\gamma_{b}\right)f^{3}_{1}+2\gamma_{f}\gamma_{b}\gamma_{0}\left(3+\gamma_{f}\right)f_{1}f_{2}+\gamma_{b}\gamma^{2}_{f}\left(2+\gamma_{b}\right)f_{3} ,
\label{pnsmall}
\end{align}
where we used for brevity $f_{n}\equiv f(nk)$. 

Physically, $\gamma_{0}$ is the probability that the particle performs an uncorrelated hop. In real space, we get
\begin{align}
p_{n}\left(\vec{r}\right) = \frac{1}{\left(2\pi\right)^{d}} \int d\vec{k}  e^{-i\vec{k}\cdot\vec{r}} \tilde{p}_{n}\left(\vec{k}\right) = \frac{\Omega_{d}}{\left(2\pi\right)^{d}} \int dk  \tilde{p}_{n}\left(\vec{k}\right) k^{d-1} sinc(kr)  .\label{pnreal}
\end{align}
Note that $\tilde{p}_{n}\left(\vec{k}\right)$ depends only on $k$ and $p_{n}\left(\vec{r}\right)$ depends only on $r$, however we write their arguments as $\vec{k}$ and $\vec{r}$ to emphasize that these are probability densities with respect to the vectors $\vec{k}$ and $\vec{r}$. 
For $n=0,1$ we get
\begin{align}
&p_{0}\left(\vec{r}\right)=\frac{\delta(r)}{\Omega_{d} r^{d-1}} ,\nonumber\\
&p_{1}\left(\vec{r}\right)=\frac{s(r)}{\Omega_{d} r^{d-1}} .\label{p01}
\end{align}
These results can also be obtained by noting that for $n=0$ the particle does not move, and thus its displacement must be zero, while for $n=1$ it performs one move with a step size distribution $s\left(\ell\right)$.

Since every time the particle performs a backward move it effectively cancels its previous hop, the probability that the particle moved a distance $\vec{r}$ given that it performed $n$ hops, $n_{b}$ of which were backwards, is propotional to the probability that it moved a distance $\vec{r}$ given that it performed $n-2n_{b}$ hops without backward hops, and thus $\tilde{p}_{n}\left(\vec{k}\right)$ for a general value of $\gamma_{b}$ may be written as a linear combination of $\tilde{p}_{n}\left(\vec{k}\right)$ for $\gamma_{b}=0$, 
\begin{align}
\tilde{p}_{n}\left(\vec{k},\gamma_{b}\right)=\sum^{\left\lfloor\frac{n}{2}\right\rfloor}_{m=0}A_{m,n}\left(\gamma_{b}\right)\tilde{p}_{n-2m}\left(\vec{k},0\right) ,
\end{align}
where $A_{n,m}(\gamma_{b})$ are some constants. Trivially, for $\gamma_{b}=0$ we have $A_{m,n}=\delta_{m,0}$. Therefore, by linearity, we find that Eq. (\ref{pdsum}) for a finite value of $\gamma_{b}$ may be written as
\begin{align}
P_{d}\left(\vec{r},\tau,\gamma_{b}\right)=\sum^{\infty}_{n=0}\tilde{q}_{n}\left(\tau\right)p_{n}\left(\vec{r},\gamma_{b}=0\right) ,
\end{align}
where $\tilde{q}_{n}$ are linear combinations of the original $q_{n}$'s. This means that a system with a finite $\gamma_{b}$ behaves the same as a system with $\gamma_{b}=0$ but with different time correlations. Therefore, for general temporal correlations it is sufficient to discuss only the case $\gamma_{b}=0$. In the limit $\gamma_{b}=0$ we can find an explicit expression for Eq. (\ref{pngen})
\begin{align}
\tilde{p}_{n}\left(\vec{k}\right) &=
\left\langle 1\right|\left[\gamma_0 {\cal M}_{0}+\gamma_{f}{\cal M}_{f}\right]^{n-1}\left|\tilde{p}_{1}\left(\vec{k}\right)\right\rangle = \nonumber\\
&=\sum^{n}_{M=1}\gamma_0^{M-1}\gamma^{n-M}_{f}\sum^{n}_{m_{1}=1}...\sum^{n}_{m_{M}=1}\delta_{n,\sum^{M}_{i=1}m_{i}}\prod^{M}_{i=1}f\left(m_{i}k\right) .
\end{align}

In conclusion, the required stages to evaluate the PDF of the displacements, $P_{r}$, in our model for a given distribution of step length $s\left(\ell\right)$ and temporal correlations encoded in $q_{n}$, are: 1) calculate the function $f(k)$ using Eq. (\ref{fdef}), 2) calculate the Fourier transform of the discrete distribution given $n$ steps $\tilde{p}_{n}\left(\vec{k}\right)$ using Eq. (\ref{pngen}), 3) evaluate the Fourier transform of the discrete distribution $P_{d}\left(\vec{k}\right)$ using the Fourier transform of Eq. (\ref{pdsum}), and 4) calculate the convolution of the discrete and continuous processes using Eq. (\ref{pr_4}).

\subsection{Specific distributions}

In this section we show how $P_{r}$ behaves for several specific choices of the step size distribution and of the temporal and orientational correlations. We consider three types of step size distributions, $s(\ell)$, and three types of temporal correlations, encoded in $q_{n}$, and examine all nine combinations. The three different distributions that we consider for the step size are: 1) a Dirac delta distribution
\begin{align}
s_{D}\left(\ell\right)=\delta\left(\ell-a\right) ,\label{s_dist_delta}
\end{align}
2) a modified Gaussian distribution
\begin{align}
s_{G}\left(\ell\right)=\left(\frac{\ell}{a}\right)^{2\nu}e^{-\nu\ell^{2}/a^{2}}\frac{2\nu^{\nu+1/2}}{a\Gamma\left(\nu+\frac{1}{2}\right)} ,\label{s_dist_gauss}
\end{align}
which in the limit $\nu\rightarrow\infty$ converges to the Dirac distribution, and 3) a Cauchy distribution
\begin{align}
s_{C}\left(\ell\right)=\frac{4a\ell^{2}}{\pi\left(\ell^{2}+a^{2}\right)^{2}} .\label{s_dist_cauchy}
\end{align}
For all three distributions the most likely step size is $a$. The mean step size for the three distributions is
\begin{align}
&\left\langle\ell\right\rangle_{D}=\int^{\infty}_{0}\ell s_{D}\left(\ell\right)d\ell=a ,\nonumber\\
&\left\langle\ell\right\rangle_{G}=\int^{\infty}_{0}\ell s_{G}\left(\ell\right)d\ell=a\frac{\sqrt{\nu}\Gamma\left(\nu\right)}{\Gamma\left(\nu+\frac{1}{2}\right)} ,\nonumber\\
&\left\langle\ell\right\rangle_{C}=\int^{\infty}_{0}\ell s_{C}\left(\ell\right)d\ell=\infty ,\label{means}
\end{align}
and the variance of the step size for the three distributions is
\begin{align}
&\sigma^{2}_{D}=\left\langle\ell^{2}\right\rangle_{D}-\left\langle\ell\right\rangle_{D}^{2}=0 ,\nonumber\\
&\sigma^{2}_{G}=\left\langle\ell^{2}\right\rangle_{G}-\left\langle\ell\right\rangle_{G}^{2}=a^{2}\left(1+\frac{1}{2\nu}-\frac{\nu\Gamma^{2}\left(\nu\right)}{\Gamma^{2}\left(\nu+\frac{1}{2}\right)}\right) ,\nonumber\\
&\sigma^{2}_{C}=\left\langle\ell^{2}\right\rangle_{C}-\left\langle\ell\right\rangle_{C}^{2}=\infty .
\end{align}
We note that for large $\nu$ the asymptotic expansions of the prefactors for the mean and variance of the modified Gaussian distribution are given by
\begin{align}
&\frac{\sqrt{\nu}\Gamma\left(\nu\right)}{\Gamma\left(\nu+\frac{1}{2}\right)}=1+\frac{1}{8\nu}+O\left(\nu^{-2}\right) ,\nonumber\\
&1+\frac{1}{2\nu}-\frac{\nu\Gamma^{2}\left(\nu\right)}{\Gamma^{2}\left(\nu+\frac{1}{2}\right)}=\frac{1}{4\nu}+O\left(\nu^{-2}\right) .\label{asymp}
\end{align}

\begin{figure}
\includegraphics[width=0.45\columnwidth]{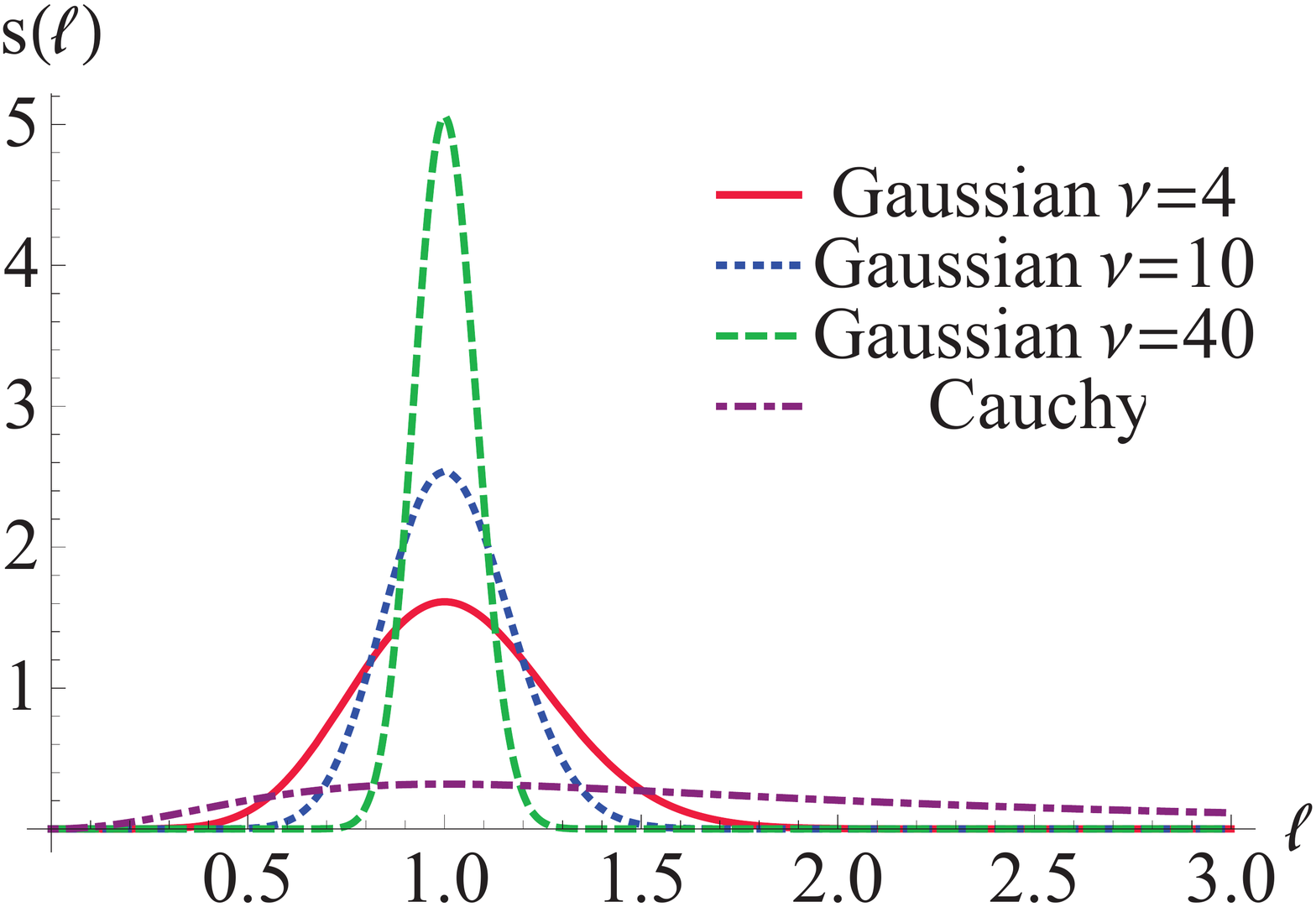}
\includegraphics[width=0.45\columnwidth]{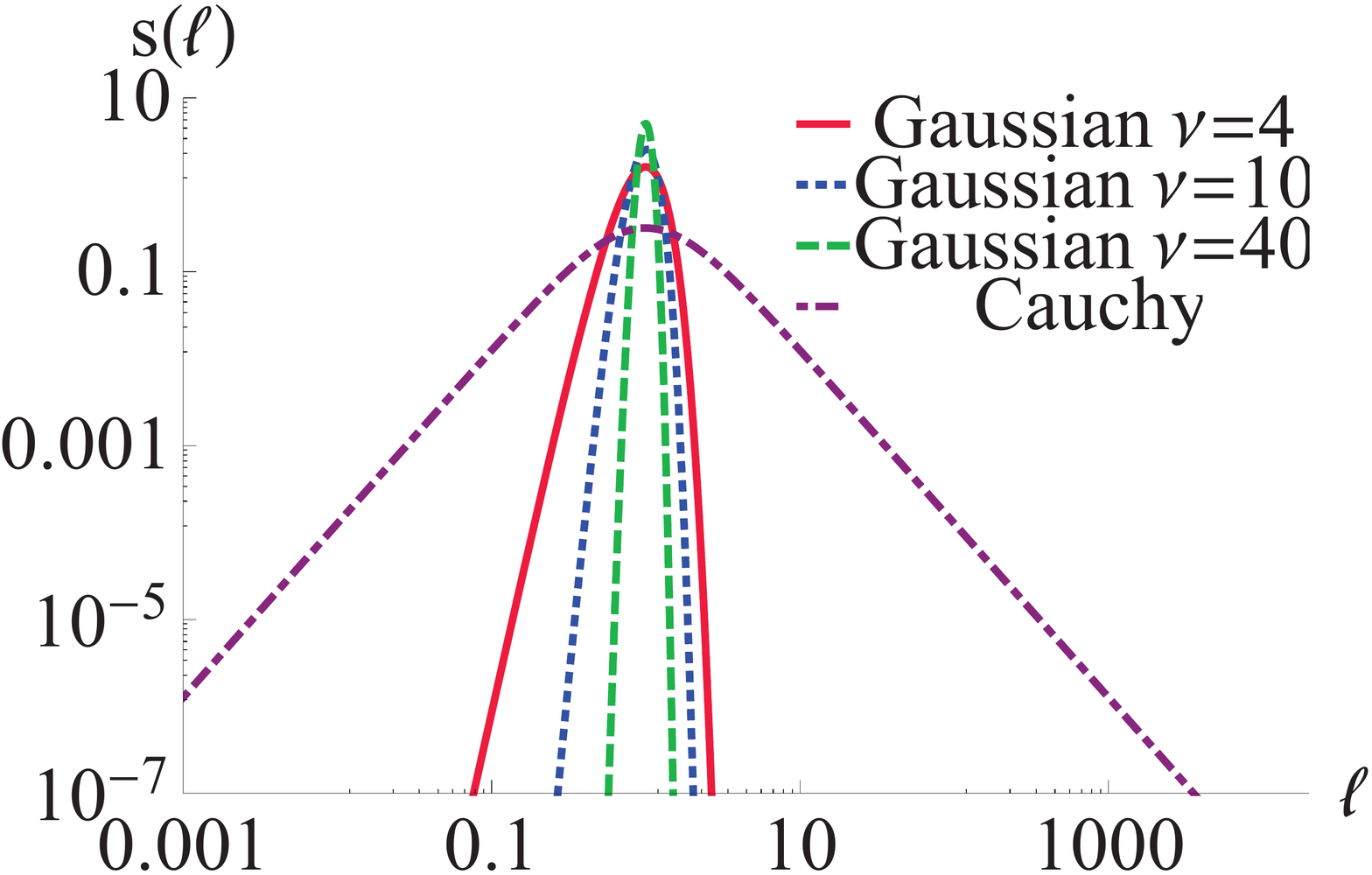}
\caption{The step size distribution $s(\ell)$ for the modified Gaussian Eq. (\ref{s_dist_gauss}) and the Cauchy Eq. (\ref{s_dist_cauchy}) distributions, all with most likely step size $a=1$.}
\label{s_dist_fig}
\end{figure}

The Dirac distribution is a natural simplistic choice. The modified Gaussian distribution is numerically almost indistinguishable from a shifted Gaussian with the same mean and variance, with the advantage that many of the calculations are analytically tractable. It is a representative of general distributions with a single peak and finite variance. From Eqs. (\ref{means}-\ref{asymp}) we see that for $\nu\gg1$ the mean over standard deviation of the Gaussian distribution is $\frac{\left\langle \ell\right\rangle_{G}}{\sigma_{G}}=2\sqrt{\nu}$. The modified Gaussian is a good approximation for every such distribution for which this ratio is large enough.
The Cauchy distribution is a simple representative of heavy tail distributions, which are rather common in biological processes \cite{Coding2008,Rupprecht2016,Reynolds2013,Palyulin2017,Detcheverry2017,Johnson2008,Kurihara2017,Ariel2017,Mwaffo2015}. The modified Gaussian distribution for $\nu=4,10,40$ and $a=1$, and the Cauchy distribution with $a=1$ are shown in Fig. \ref{s_dist_fig}.
The function $f(k)$ defined in Eq. (\ref{fdef}) for these three distributions is
\begin{align}
&f_{D}(k)=sinc\left(ak\right) ,\nonumber\\
&f_{G}(k)={}_{1}F_{1}\left(\nu+\frac{1}{2},\frac{3}{2},-\frac{a^{2}k^{2}}{4\nu}\right) ,\nonumber\\
&f_{C}(k)=e^{-ak} .
\end{align}

In the modified Gaussian distribution, we find that for integer values of $\nu$
\begin{align}
f_{G}(k)=\frac{\left(-1\right)^{\nu-1}\left(\nu-1\right)!\sqrt{\nu}e^{-a^{2}k^{2}/4\nu}H_{2\nu-1}\left(\frac{ak}{2\sqrt{\nu}}\right)}{ak\left(2\nu-1\right)!} ,\label{fint}
\end{align}
where $H$ are the Hermite polynomials \cite{hermite}.

For the Cauchy distribution we find that when $\gamma_{b}=0$ then
\begin{align}
\tilde{p}^{C}_{n}(k)=e^{-nak} ,
\end{align}
for any value of $\gamma_{f}$, and thus $p^{C}_{n}(r)$ is
\begin{align}
&p^{C}_{n}(r)=\frac{\left(d-2\right)!}{r\left[r^{2}+\left(na\right)^{2}\right]^{(d+1)/2}}\times\nonumber\\
&\left\{\left[\left(na\right)^{2}-r^{2}\right]\sin\left[\left(d+1\right)\cot^{-1}\left(\frac{na}{r}\right)\right]-2nar\cdot\cos\left[\left(d+1\right)\cot^{-1}\left(\frac{na}{r}\right)\right]\right\} .
\end{align}
By including the continuous process, we find that in three dimensional systems [see Eqs. (\ref{pdsum}) and (\ref{pr_4_d3})],
\begin{align}
&\hat{p}^{C}_{n}(r)=\frac{1}{2\pi^{2}r}\int^{\infty}_{0}dk e^{-nak}k\sin(kr)e^{-\frac{\alpha^{2}k^{2}}{2}}=\nonumber\\
&=\frac{e^{-\frac{r^{2}-n^{2}a^{2}}{2\alpha^{2}}}}{4\sqrt{2}i\pi^{3/2}\alpha^{3}r}\left\{\left(-na+ir\right)e^{\frac{-inar}{\alpha^{2}}}\left[1+Erf\left(\frac{-na+ir}{\sqrt{2}\alpha}\right)\right]-\right.\nonumber\\
&\left.-\left(-na-ir\right)e^{\frac{inar}{\alpha^{2}}}\left[1+Erf\left(\frac{-na-ir}{\sqrt{2}\alpha}\right)\right]\right\} ,
\end{align}
with $Erf(z)$ being the error function \cite{erf} defined by
\begin{align}
Erf(z)=\frac{2}{\sqrt{\pi}}\int^{z}_{0}e^{-t^{2}}dt .
\end{align}

The three types of temporal correlations that we consider are modeled by three possible distributions $q_{n}$: 1) Poissonian
\begin{align}
q^{Pois}_{n}\left(\tau\right)=\frac{\tau^{n}e^{-\tau}}{n!} ,
\end{align}
2) binomial
\begin{align}
q^{bin}_{n}\left(\tau\right)=\left(\begin{array}{c}\tau\\n\end{array}\right)p^{n}\left(1-p\right)^{\tau-n} ,
\end{align}
with $p=0.6$, and 3) geometric
\begin{align}
q^{geo}_{n}\left(\tau\right)=\tau^{-1}\left(1-\tau^{-1}\right)^{n} .
\end{align}
In the Poissonian distribution there are no temporal correlations between the hops. The main qualitative difference between the binomial and the geometric distribution is that for the geometric distribution $\frac{\partial q^{geo}_{n}}{\partial n}<0$ always, i.e. more steps are always less likely, while in the binomial distribution the most likely number of steps is finite. Physically, the geometric distribution corresponds to the case where the waiting time between hops grows with time, while the binomial distribution corresponds to the case where only a finite maximal number of steps is allowed at any time interval. In all cases we chose the parameters so that on average the walker performs three discrete steps during the time interval $\tau$, i.e.
\begin{align}
\left\langle n\right\rangle=\sum^{\infty}_{n=0}nq_{n}=3. 
\end{align}

\subsection{Numerical Results}

\begin{figure}
\includegraphics[width=0.6\columnwidth]{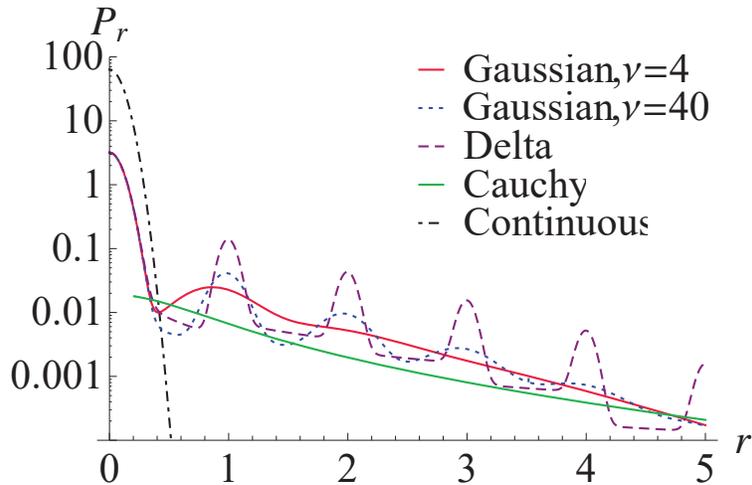}
\caption{The PDF of the walker's displacement $P(r)$ as a function of $r$ for different distributions in three dimensional systems with a Poissonian distribution of $q_{n}$. We set $\gamma_{b}=0$, $a=1$ and $\alpha=0.1$ throughout.}
\label{pr_dist}
\end{figure}

We evaluated all the expressions above in order to get $P(r)$. We start by comparing the different step size distributions without temporal correlations. The corresponding results are shown in Fig. \ref{pr_dist}. As expected, the Cauchy distribution shows no peaks in $P_{r}(r)$. For the Gaussian and the Dirac distributions, we find that the peaks, when they are seen, are located at integer multiples of $a$ and are more pronounced as the step size distribution is more narrow.

\begin{figure}
\includegraphics[width=0.3\columnwidth]{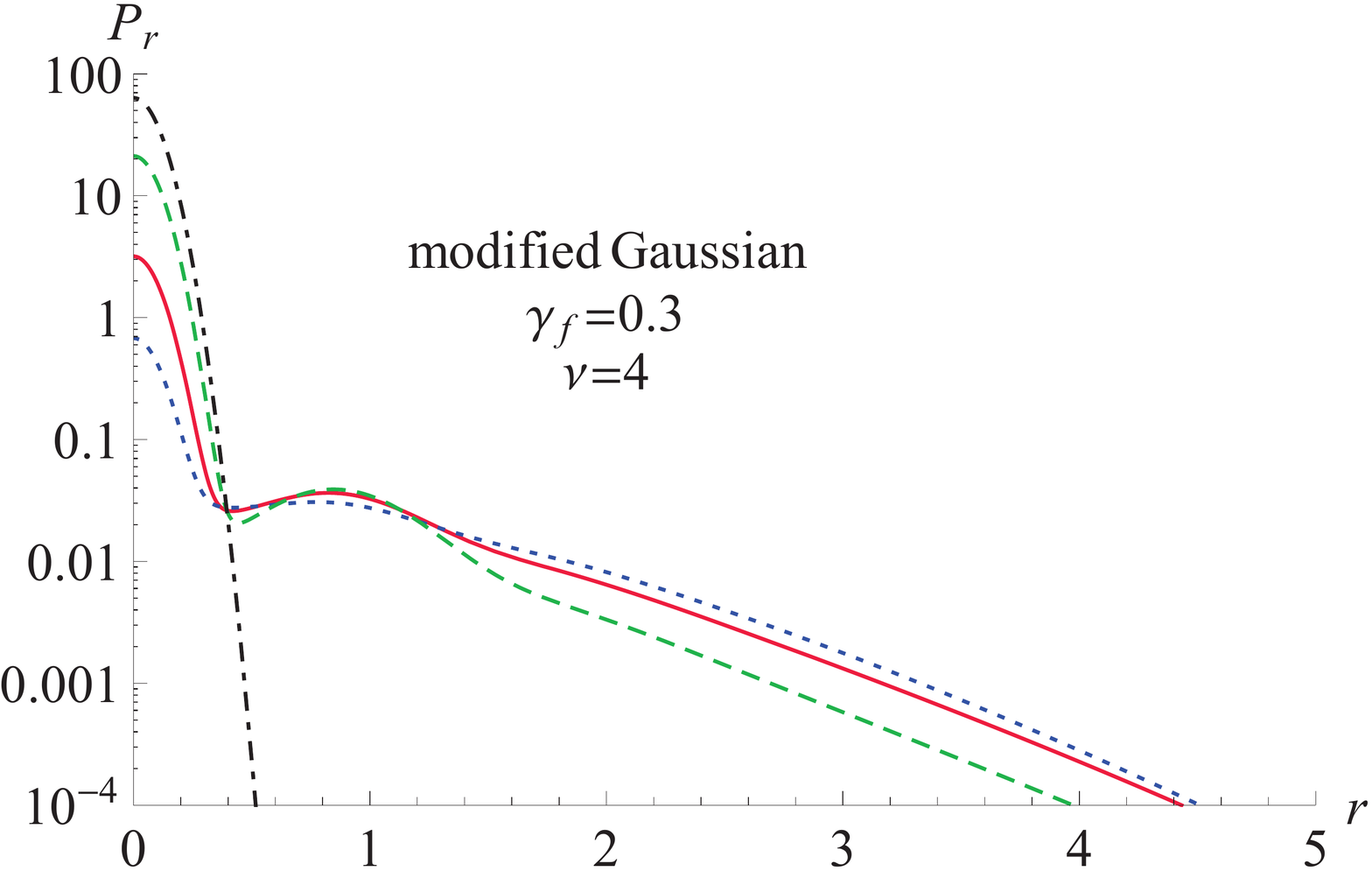}
\includegraphics[width=0.3\columnwidth]{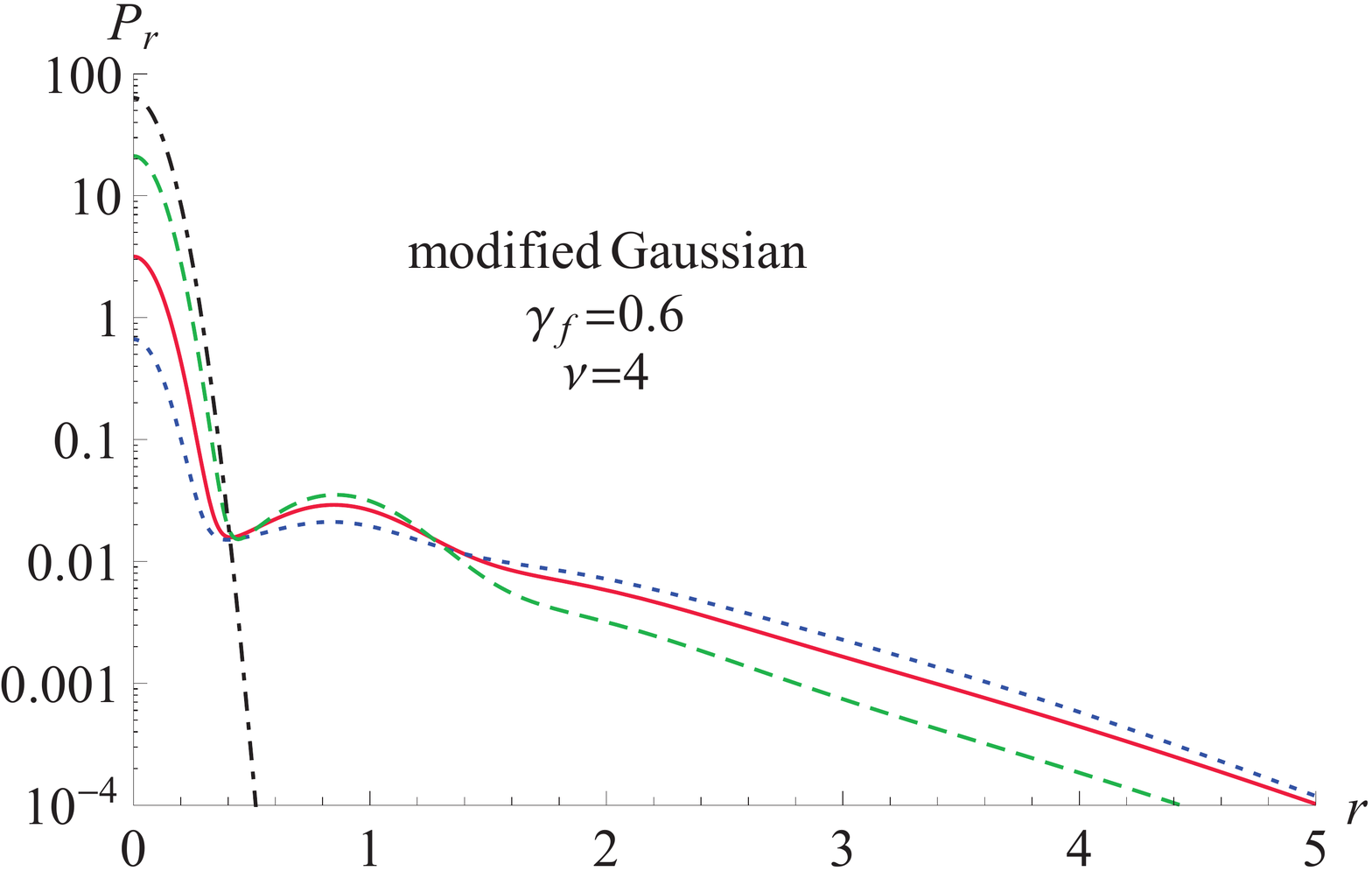}
\includegraphics[width=0.3\columnwidth]{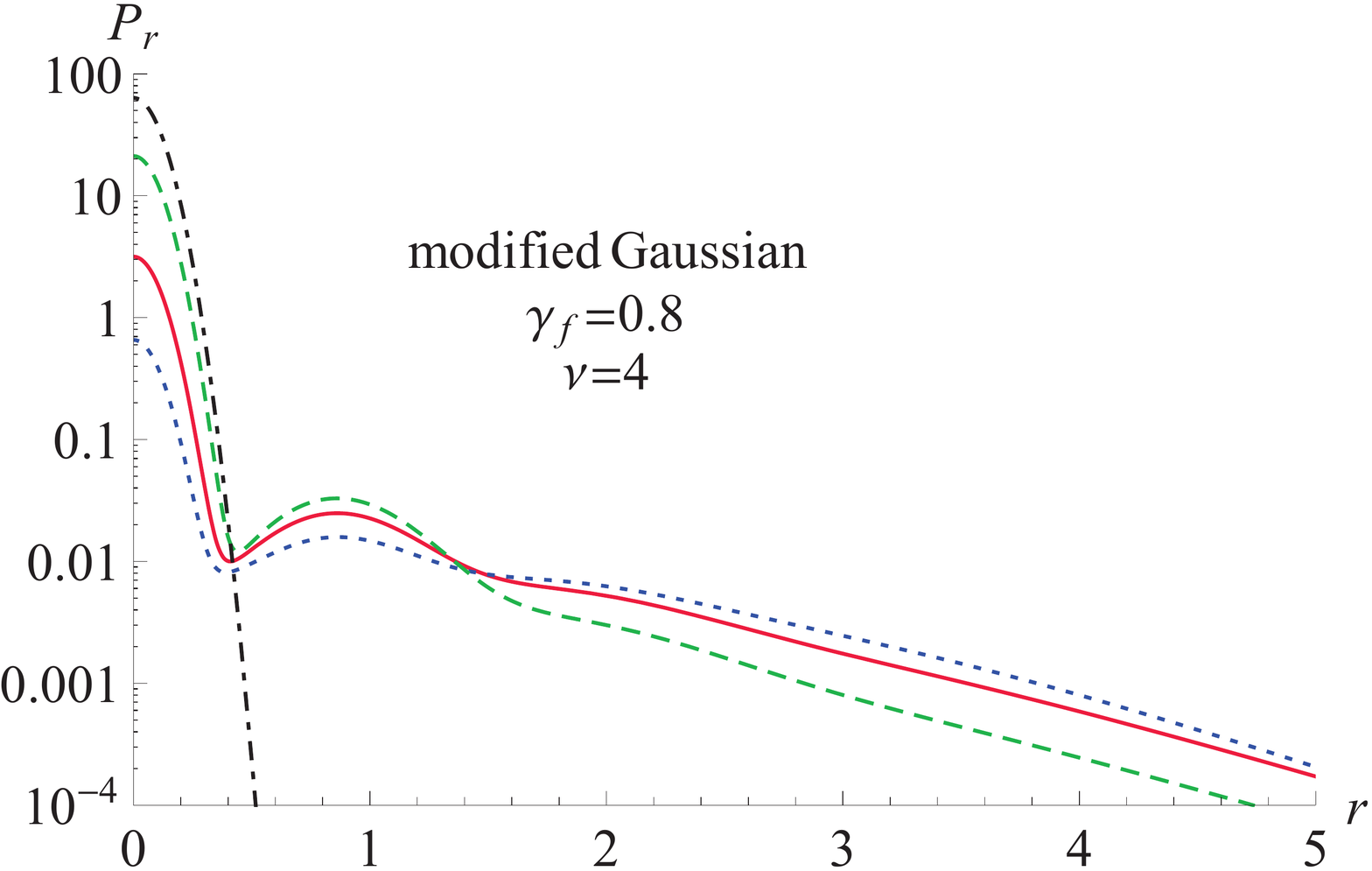}\\
\includegraphics[width=0.3\columnwidth]{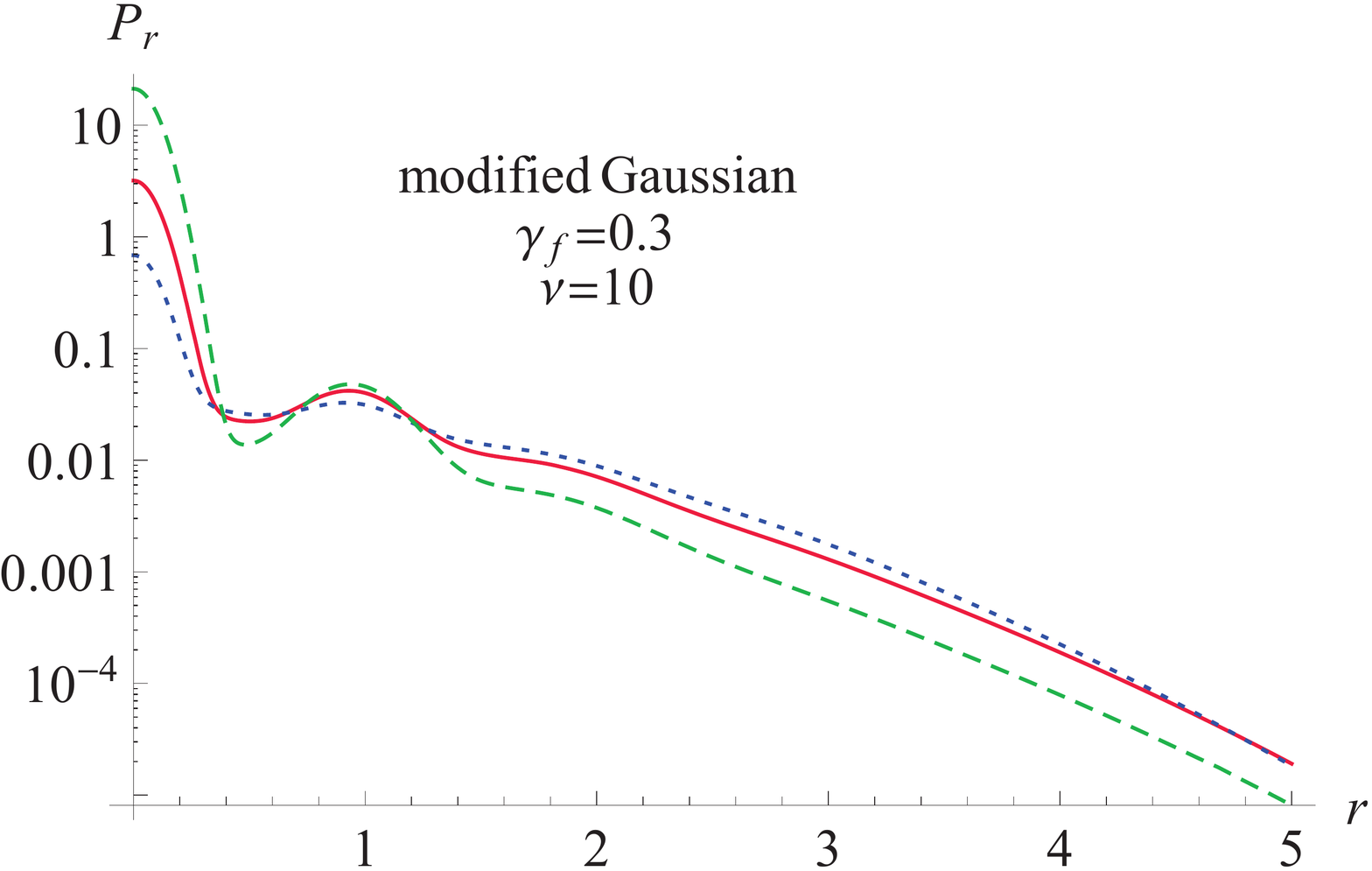}
\includegraphics[width=0.3\columnwidth]{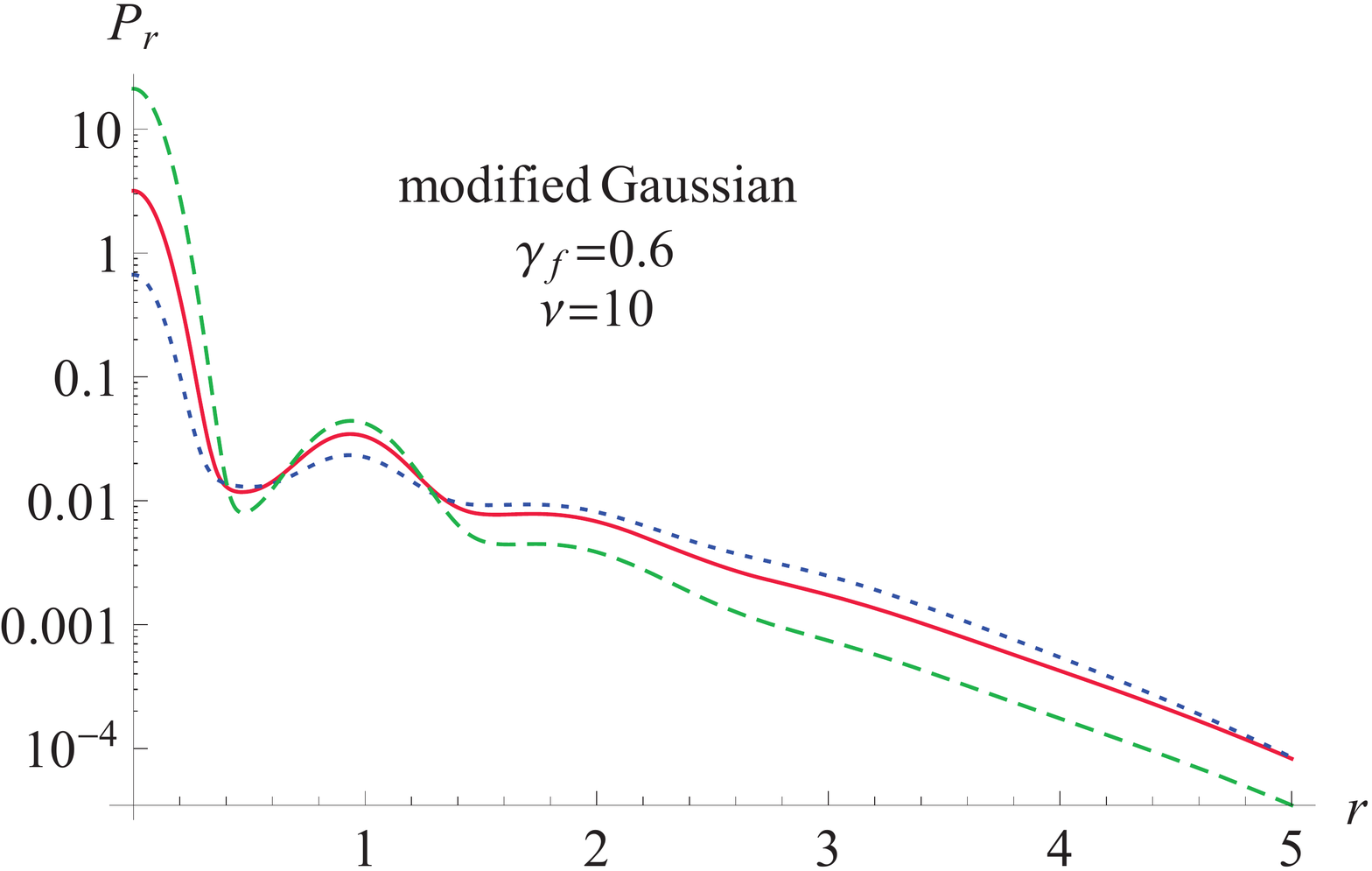}
\includegraphics[width=0.3\columnwidth]{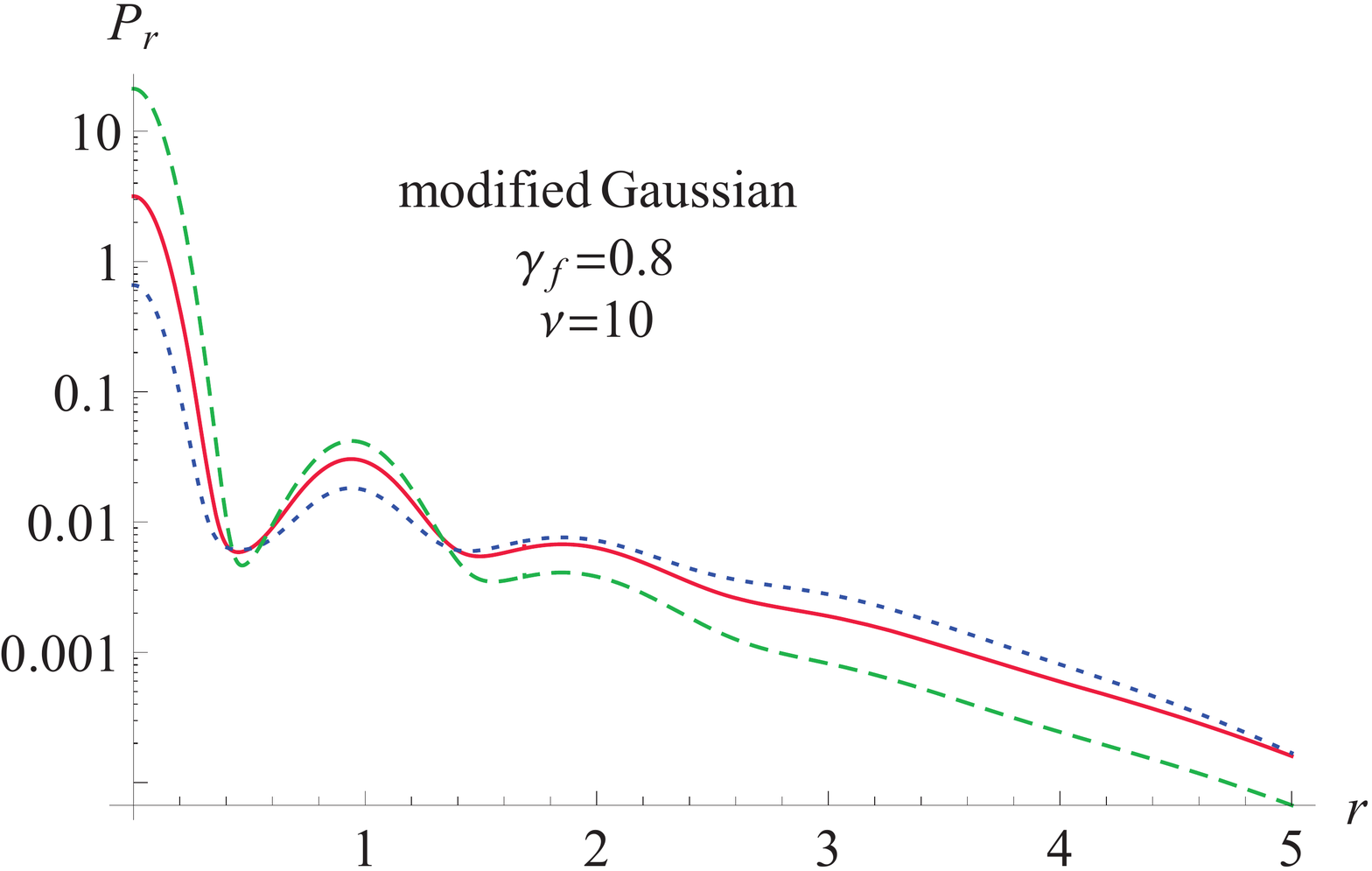}\\
\includegraphics[width=0.3\columnwidth]{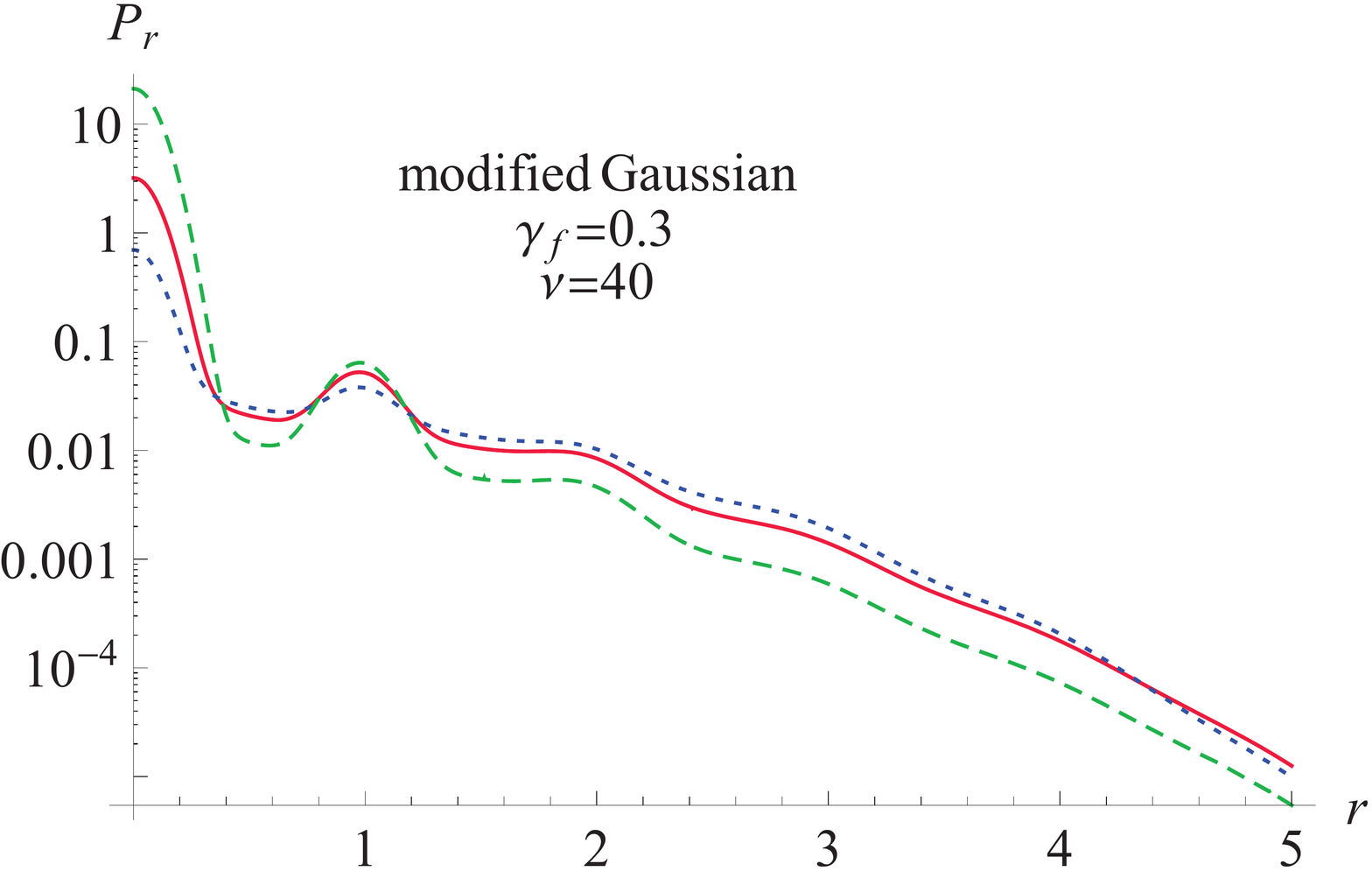}
\includegraphics[width=0.3\columnwidth]{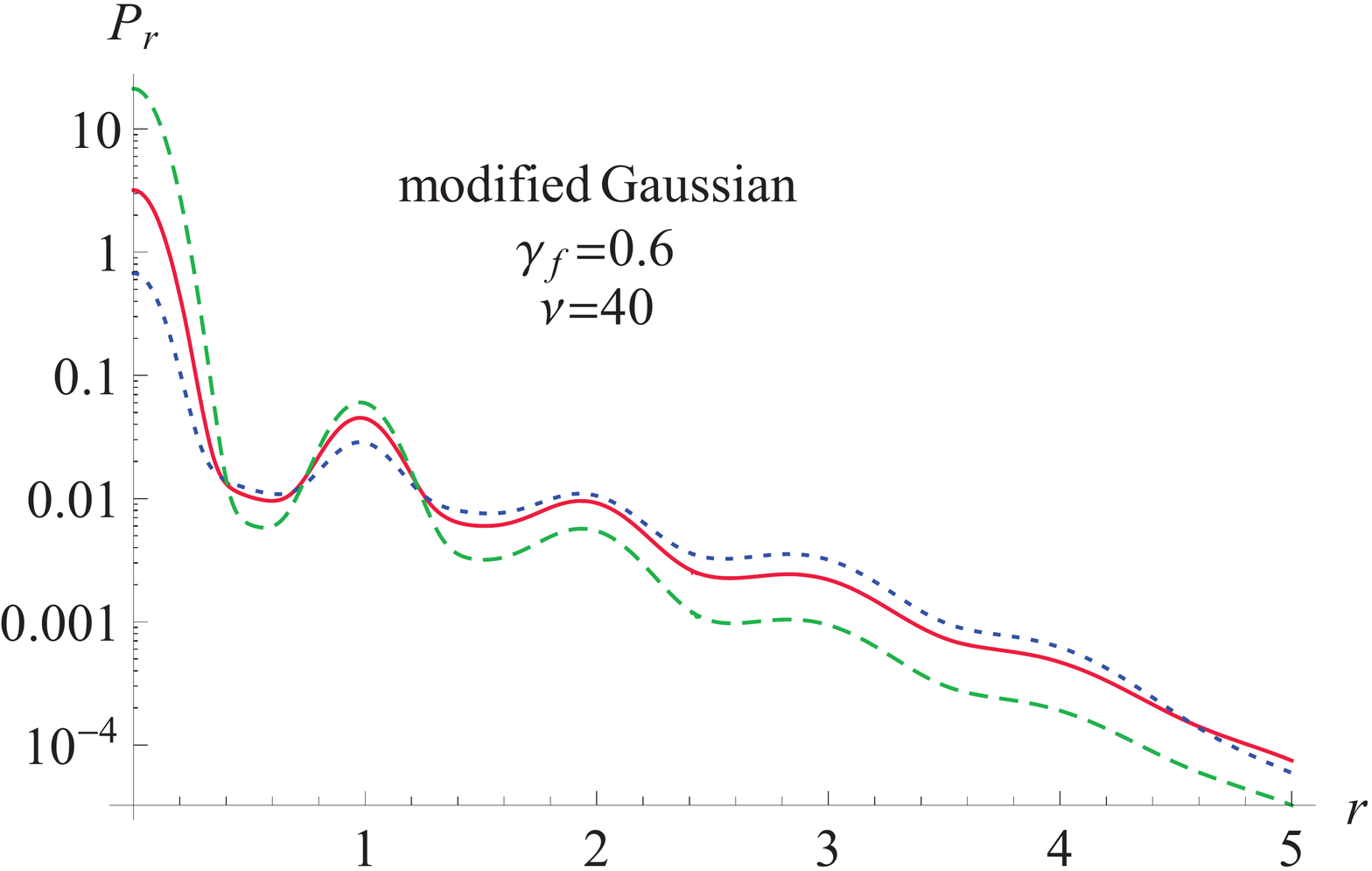}
\includegraphics[width=0.3\columnwidth]{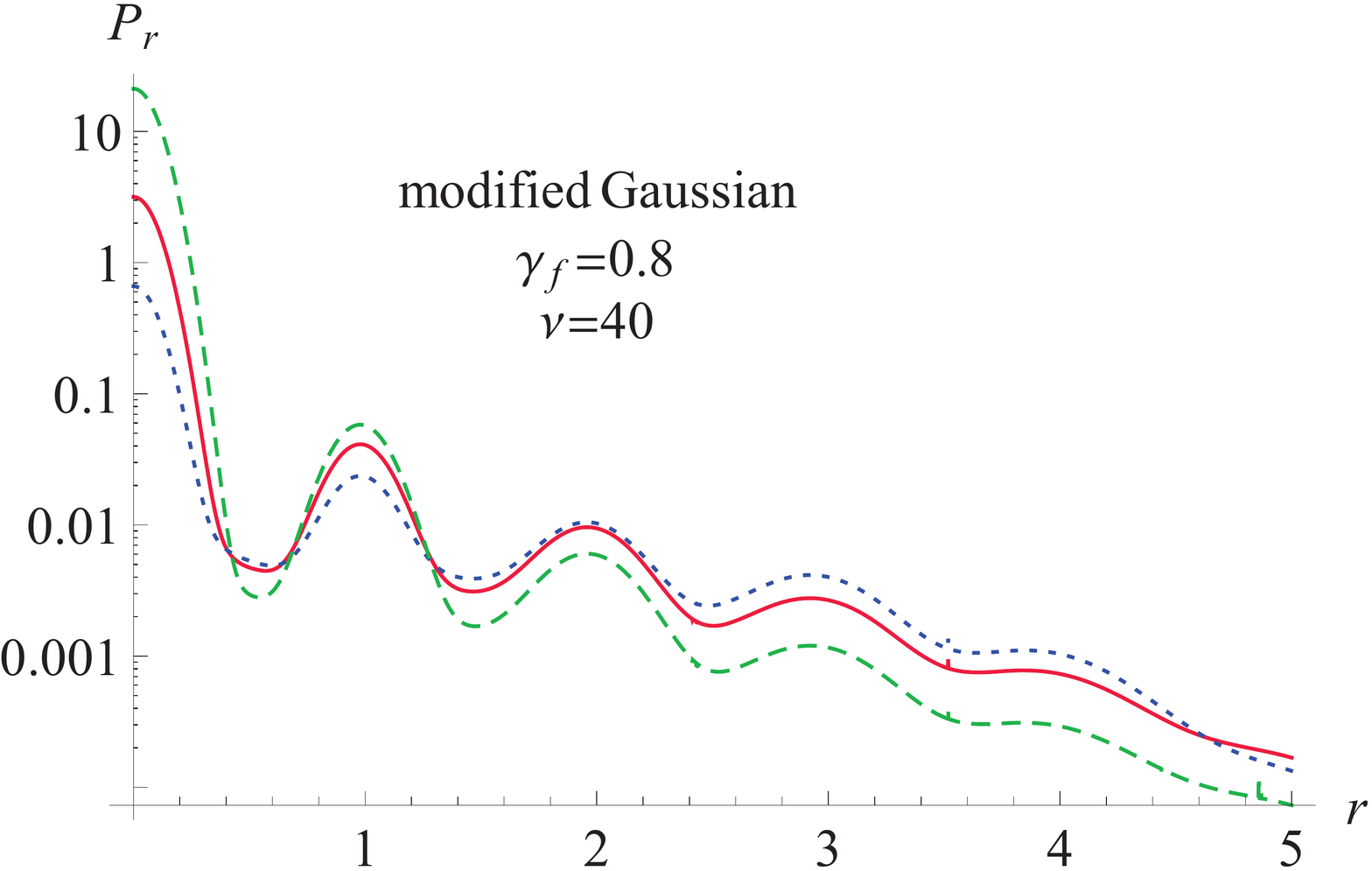}\\
\includegraphics[width=0.3\columnwidth]{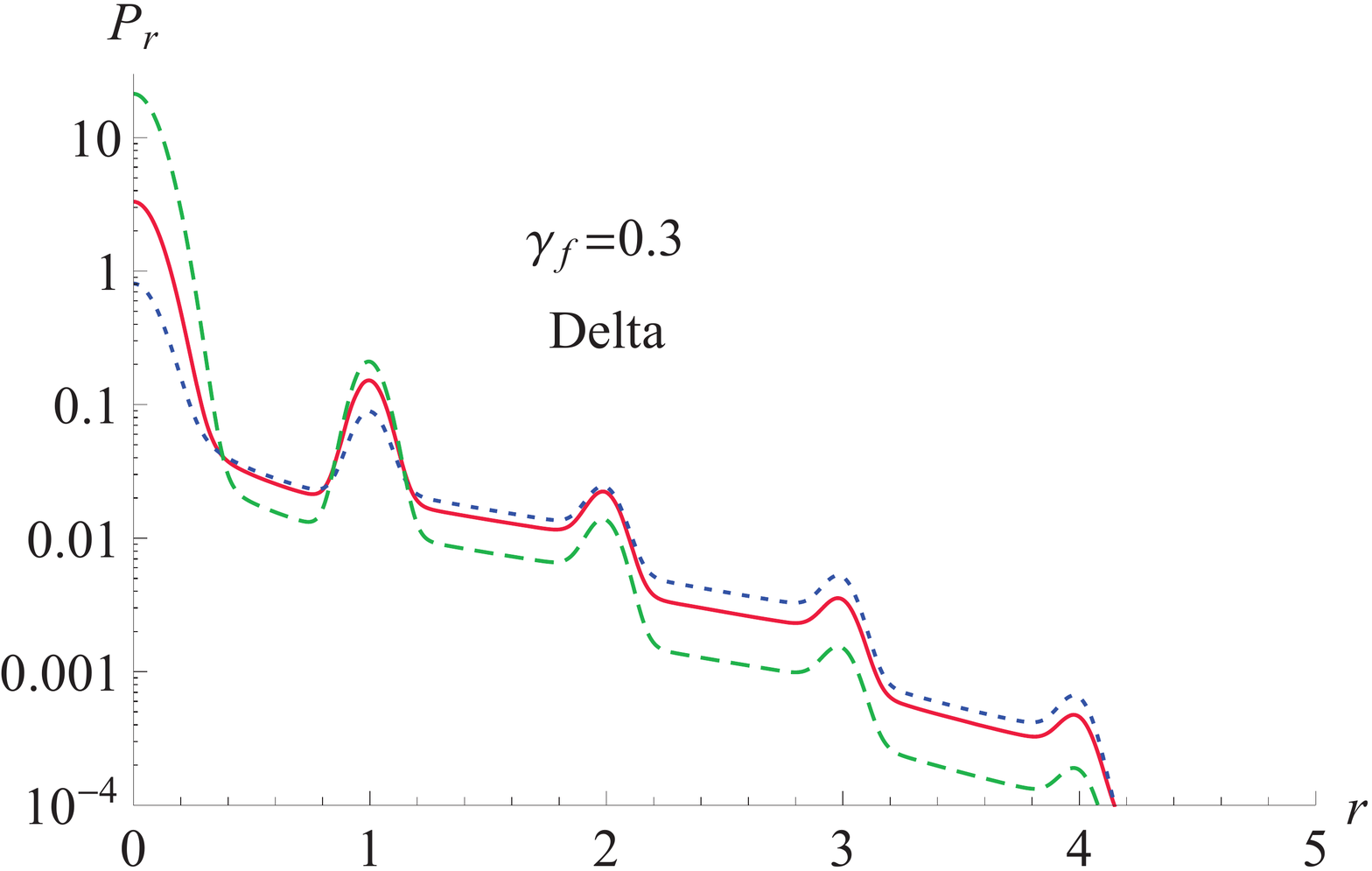}
\includegraphics[width=0.3\columnwidth]{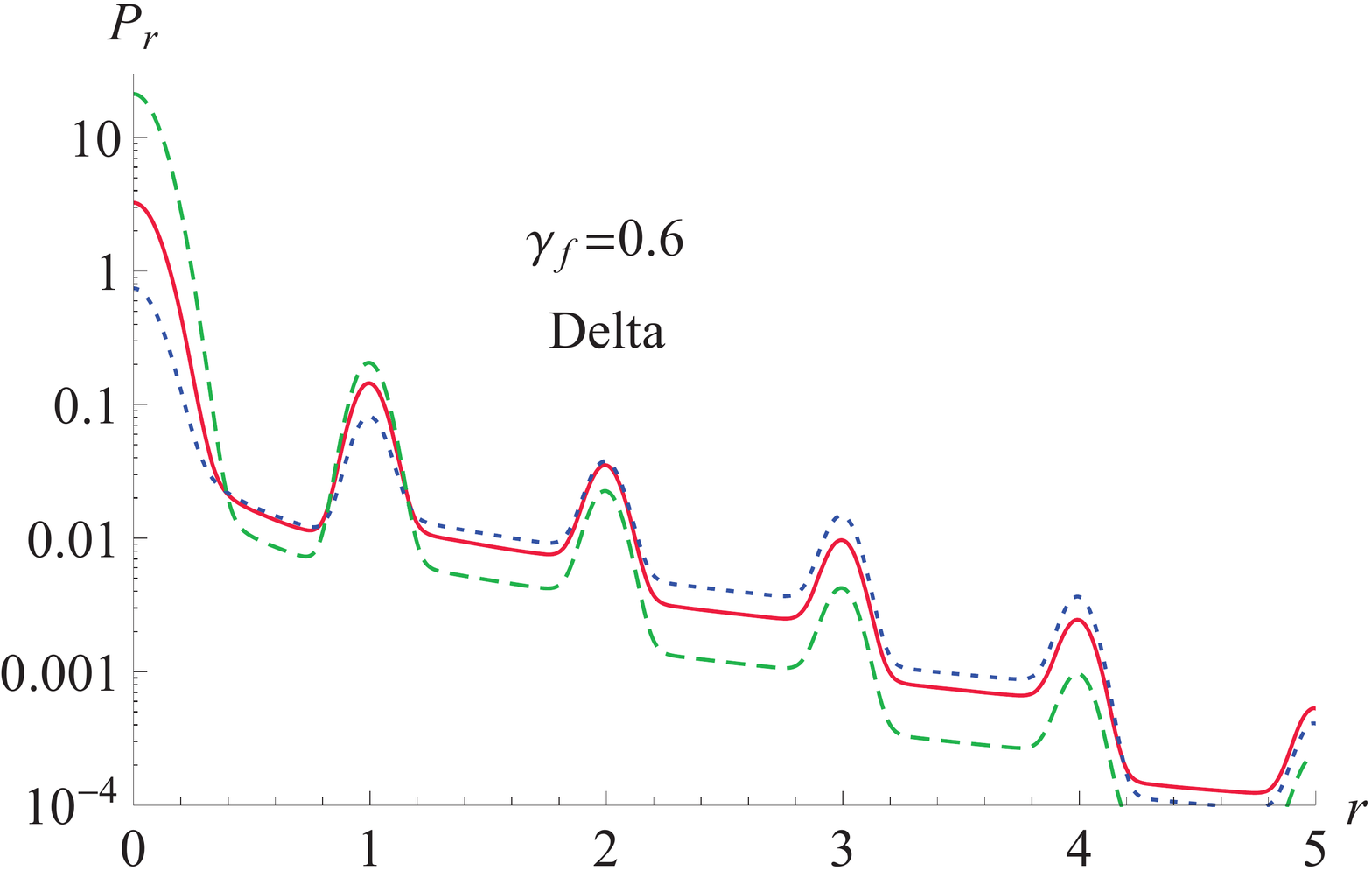}
\includegraphics[width=0.3\columnwidth]{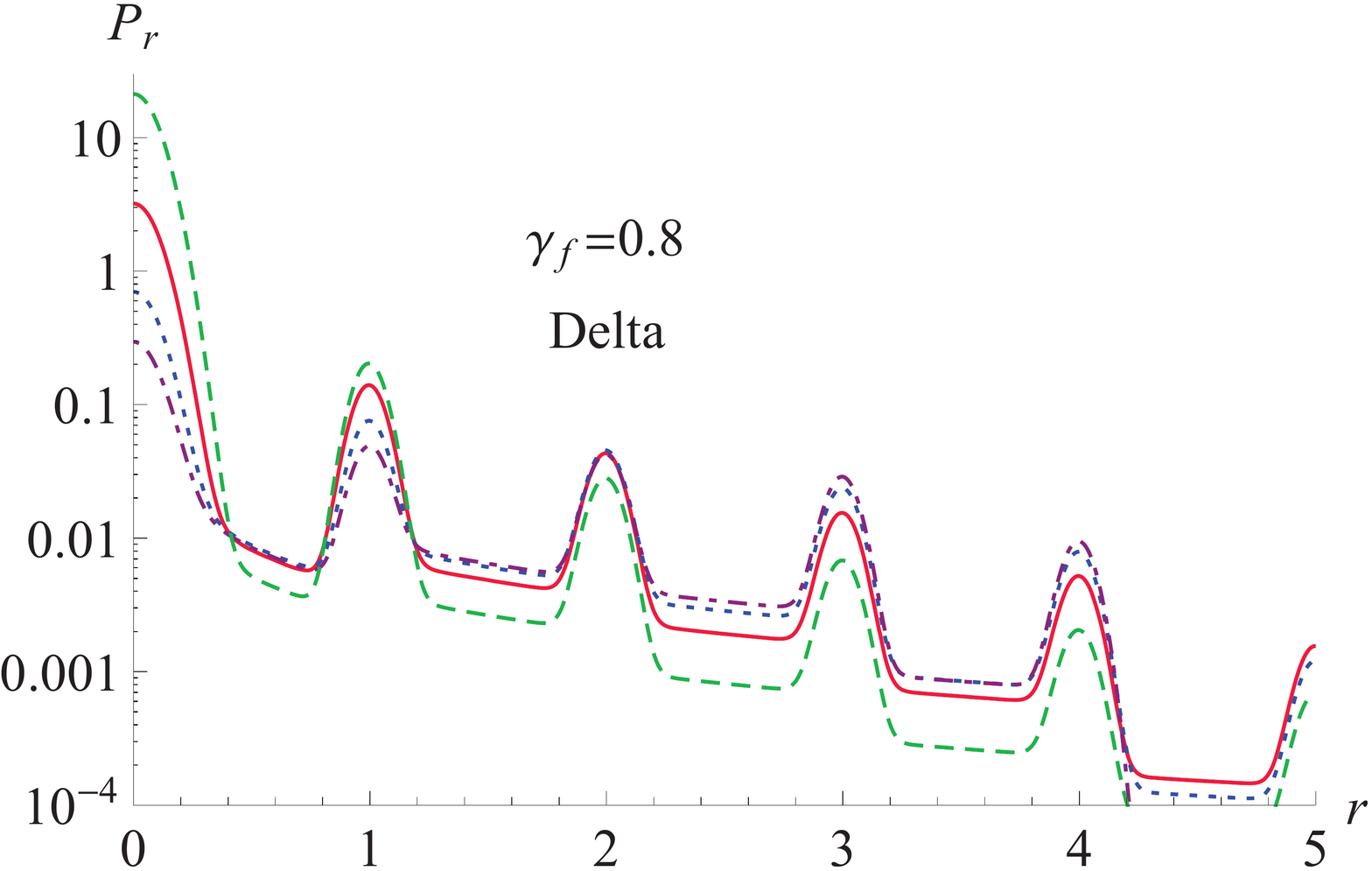}\\
\includegraphics[width=0.3\columnwidth]{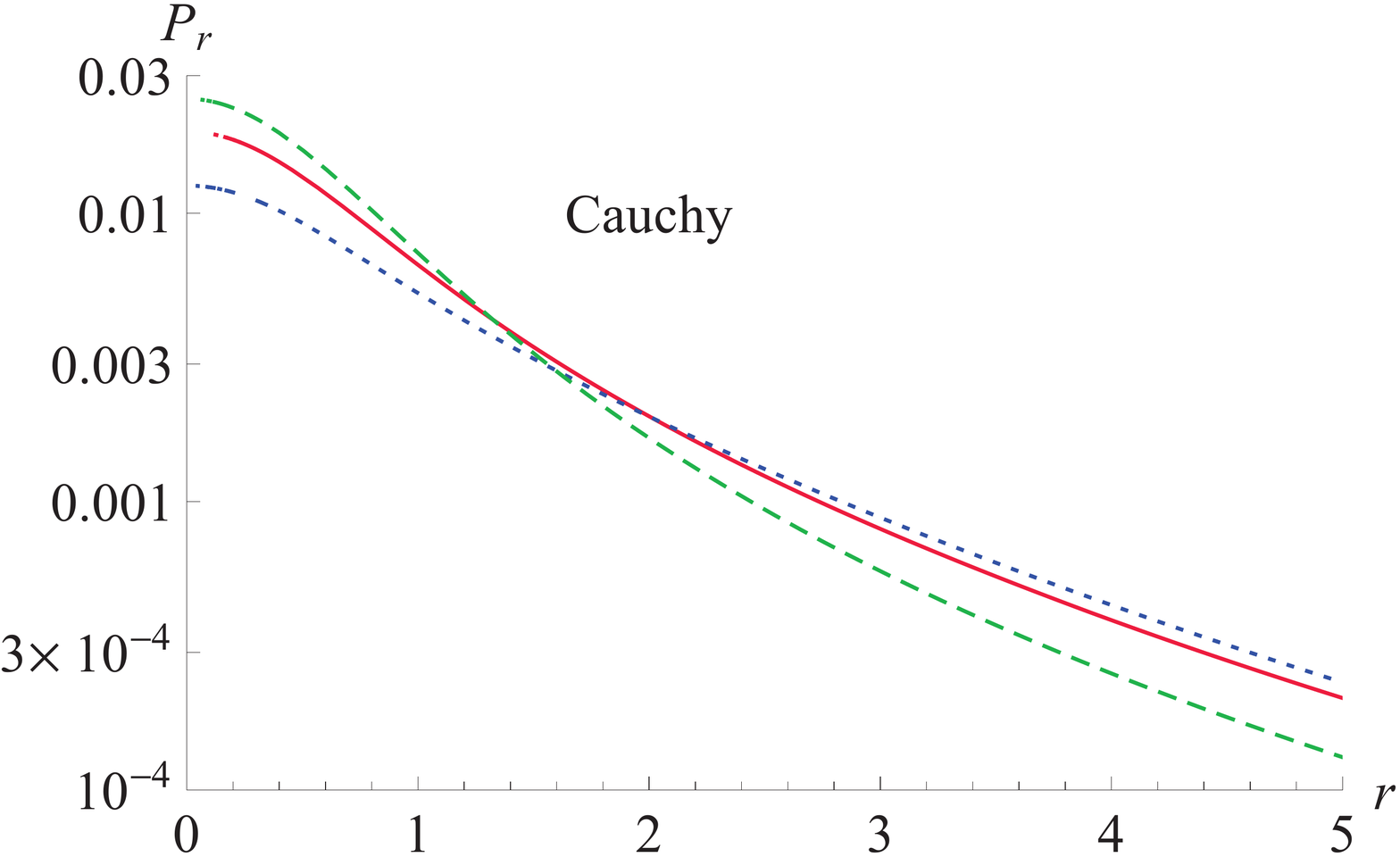}
\includegraphics[width=0.3\columnwidth]{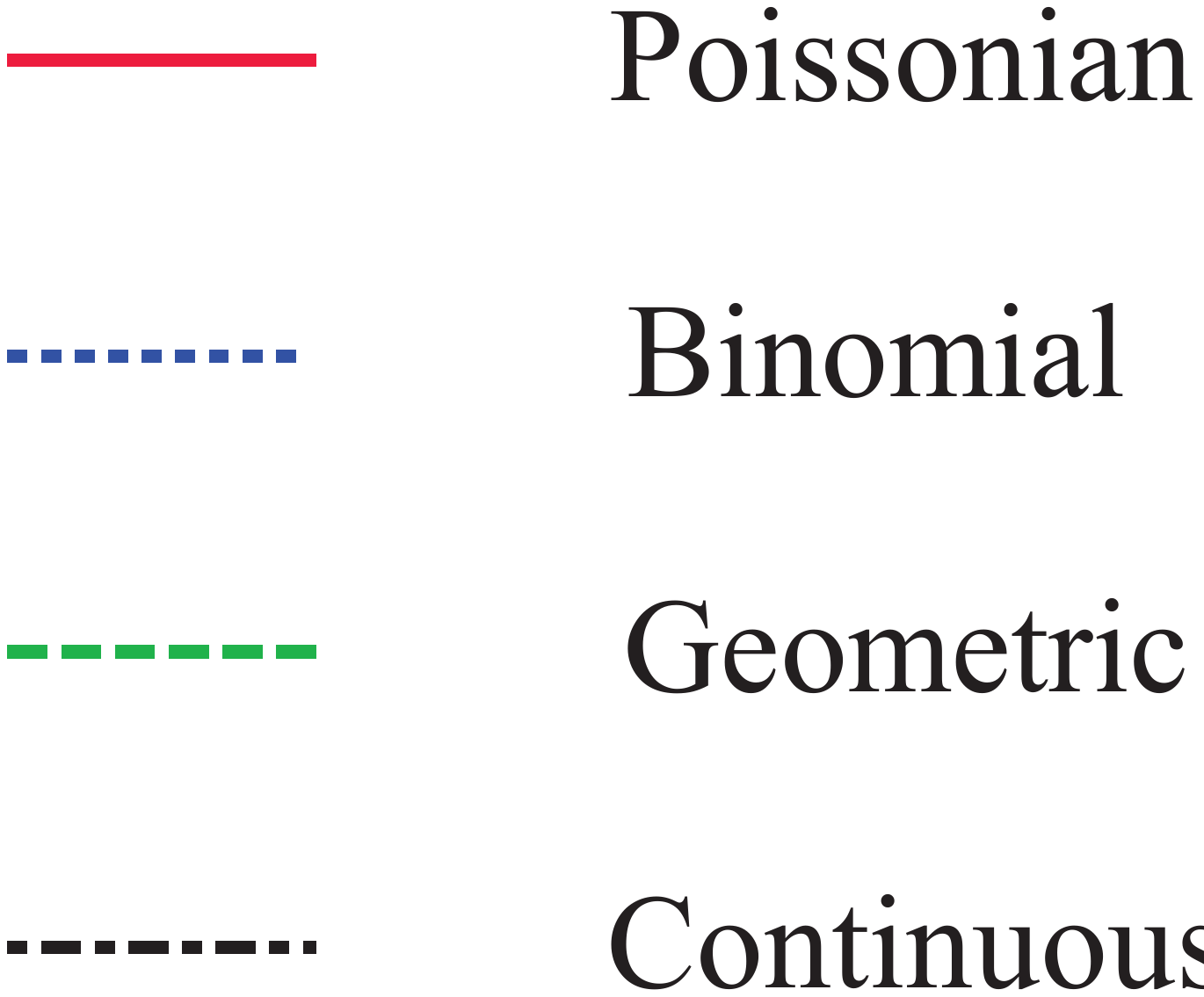}
\caption{The PDF of the walker's displacement $P(r)$ as a function of $r$ for different distributions in three dimensional systems. The first three rows correspond to a modified Gaussian distributed step size, Eq. (\ref{s_dist_gauss}), with different values of $\nu$ and $\gamma_{f}$ as noted in each panel, the fourth row shows the results for the Dirac delta step size distribution, Eq. (\ref{s_dist_delta}), with different values of $\gamma_{f}$, and the last panel shows the results for the Cauchy distribution, Eq. (\ref{s_dist_cauchy}), for which the value of $\gamma_{f}$ is immaterial. Each line corresponds to a different temporal distribution according to the legend. The dot-dashed black line, which appears only in the top row, is the displacement distribution of the continuous process alone, Eq. (\ref{pc_def}). We set $\gamma_{b}=0$, $a=1$ and $\alpha=0.1$ throughout.}
\label{dist_fig}
\end{figure}

\begin{figure}
\includegraphics[width=0.5\columnwidth]{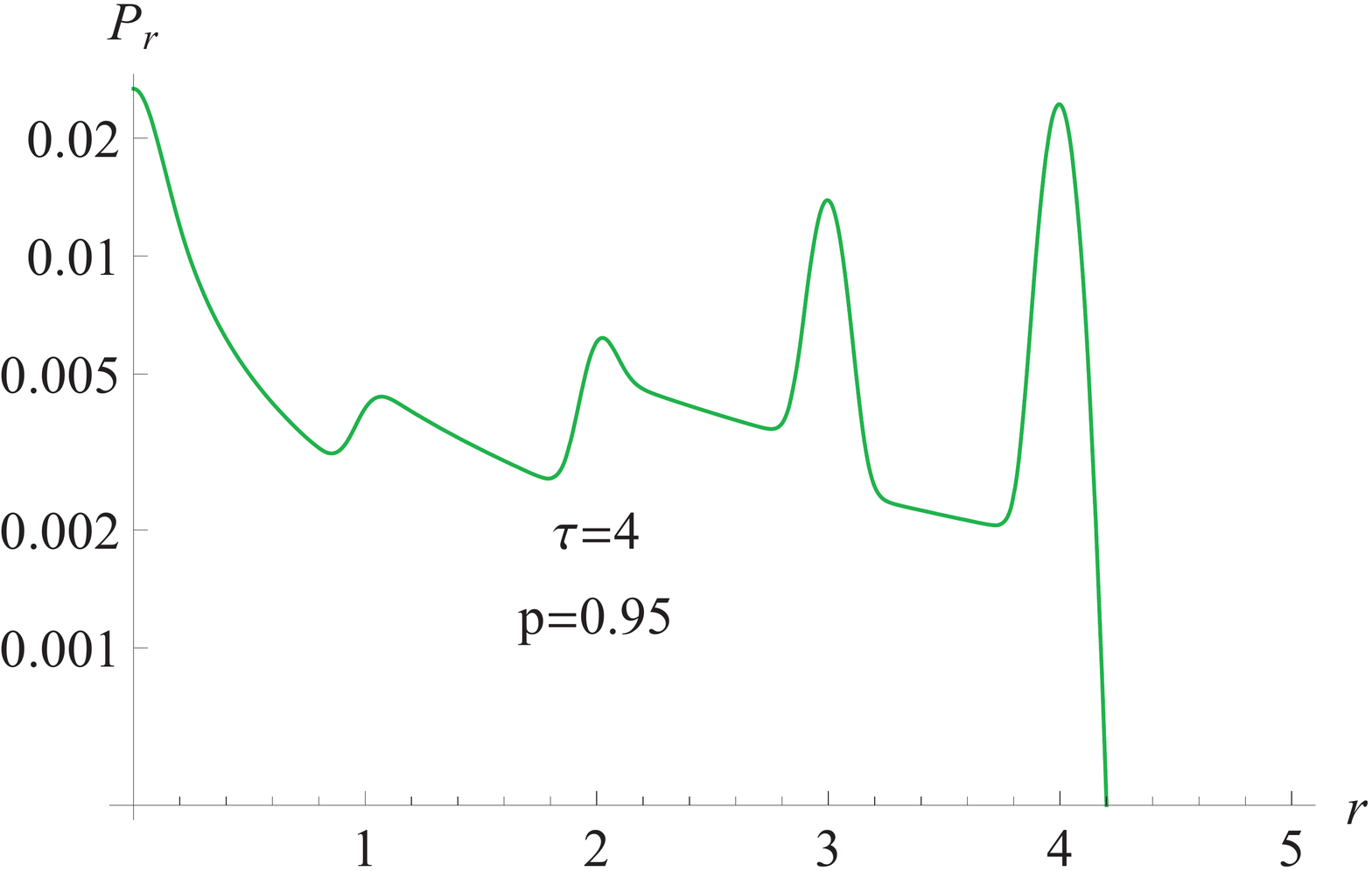}
\caption{The distribution $P_{r}(r)$ as a function of $r$ for the binomial distribution at $\tau=4$ with $p=0.95$ and a Dirac delta step size distribution. The other parameters are $a=1,\alpha=0.1,\gamma_{b}=0,\gamma_{f}=0.8$.}
\label{bin_fig}
\end{figure}

We now consider the effect of different temporal correlations and of varying the persistence $\gamma_{f}$, see Fig. \ref{dist_fig}. We find that the peaks are more pronounced as the probability to move forward $\gamma_{f}$ is larger.
For the Poissonian waiting time distribution we find that the height of the peaks decreases exponentially with the displacement, while for the other distributions it does not even necessarily decrease, as shown in the example plotted in Fig. \ref{bin_fig}.

In conclusion, if peaks are found in a PDF, one should look at several factors. The distance between peaks is the mean size of the discrete step. The height of the peaks correlate with higher values of $\gamma_{f}$. The width of the peak at $r=0$ is mostly due to the continuous process. The width of the peaks is determined by a combination of the continuous process and the variance of the discrete step size distribution. Except for the Dirac delta step size distribution, the width of the peaks increases for larger displacements, and thus only a finite number of peaks is visible. The relative height of the peaks gives information about the temporal correlations. If the heights decrease exponentially with the displacement, then there are no temporal correlations. A faster-than-exponential decrease in the relative height, as in the geometric distribution, is related to negative temporal correlations, i.e. tendency to have fewer moves. Extreme cases, such as that shown in Fig. \ref{bin_fig} are related to a higher probability of having many moves.

\section{No peaks without orientational correlations}
\label{sec_proof}

Here we will show that $P_{r}\left(r,\tau\right)$ has at most one peak if there are no positive orientational correlations, i.e. $\gamma_{f}=0$. If $\gamma_{f}=0$, then by Eqs. (\ref{pngen}) and (\ref{pnsmall}) we find that $\tilde{p}_{n}(k)$ may be written as
\begin{align}
\tilde{p}_{n}(k)=\sum^{n}_{m=0}c_{m}\tilde{p}^{0}_{m}(k) ,
\end{align}
where $\tilde{p}^{0}_{m}(k)$ is the value of $\tilde{p}_{m}(k)$ at $\gamma_{b}=0$, and $c_{m}$ are non-negative coefficients whose exact value is immaterial for the arguments presented in this section.

The continuous Gaussian process, Eq. (\ref{pc_def}), can only smear the peaks, if they exist, and thus we may consider only the case $\alpha=0$. Hence, in order to show that $P_{r}(r)$ has at most one peak, it is enough to show that $p_{1}(r)$ has one peak and that $p_{n\geq2}(r)$ are non-increasing functions of $r$. Furthermore, from the same reasoning, it is enough to consider the step size distribution $s(\ell)=\delta\left(\ell-a\right)$, since for any other step size distribution peaks, if they exist, will be more smeared. 

First, let us consider $p_{1}(r)$, which using Eq. (\ref{p01}) and the Dirac distribution, is given by
\begin{align}
p_{1}(r)=\frac{\delta\left(\ell-a\right)}{\Omega_{d}r^{d-1}} .
\end{align}
It obviously has a single peak at $r=a$, and is zero elsewhere.

For $n\geq2$, we start from Eq. (\ref{pnrec}), and note that in the case $\gamma_f=0$ there is no memory, such that
\begin{align}
&p_{n+1}(r)=\int d\ell d\hat{r}p_{n+1,\ell\hat{r}}\left(\vec{r}\right) = \int d\ell d\hat{r}\frac{\delta(\ell-a)}{\Omega_{d}}\int d\ell' d\hat{r}' p_{n,\ell'\hat{r}'}\left(|\vec{r}-\ell\hat{r}|\right)=\nonumber\\
&=\int d\ell d\hat{r}\frac{\delta(\ell-a)}{\Omega_{d}} p_{n}\left(\sqrt{r^{2}+\ell^{2}-2\ell\vec{r}\cdot\hat{r}}\right) .
\end{align}
Performing the integral over $\ell$, and writing the integral over $\hat{r}$ as an integral over the $d-1$ angles yields
\begin{align}
p_{n+1}(r)=\frac{1}{\Omega_{d}}\int d\vec{\phi}_{d-2}\sin\theta d\theta p_{n}\left(\sqrt{r^{2}+a^{2}-2ar\cos\theta}\right) ,
\end{align}
where $d\vec{\phi}_{d-2}$ contains all the other $d-2$ angles except for $\theta$.
Performing the integral over $\vec{\phi}_{d-2}$ and changing the integration variable from $\theta$ to $r'=\sqrt{r^{2}+a^{2}-2ar\cos\theta}$ yields
\begin{align}
p_{n+1}(r)=\frac{1}{2a}\int^{r+a}_{|r-a|}\frac{r'}{r}p_{n}\left(r'\right)dr' .\label{pnrecg0}
\end{align}
We now use Eq. (\ref{pnrecg0}) twice, such that
\begin{align}
p_{n+2}(r)=\frac{1}{4a^{2}}\int^{r+a}_{|r-a|}\frac{r'}{r}dr'\int^{r'+a}_{|r'-a|}\frac{r''}{r'}p_{n}\left(r''\right)dr'' .\label{pnrecg02}
\end{align}
We now prove by induction on $n$ that $p_{n}(r)$ is a non-increasing function of $r$. For $n=2$ and $n=3$ we can explicitly calculate $p_{n}(r)$
\begin{align}
&p_{2}(r)=\frac{1}{2\Omega_{d} a^{d-1}r}\Theta\left(2a-r\right) ,\nonumber\\
&p_{3}(r)=\frac{1}{2\Omega_{d} a^{3}}\left[\Theta\left(a-r\right)+\Theta\left(r-a\right)\Theta\left(3a-r\right)\frac{3a-r}{2r}\right] ,
\end{align}
which are non-increasing functions of $r$. For $n\geq4$ we consider three different cases: $r>2a$, $2a>r>a$, and $a>r$.

\subsection{$r>2a$}
If $r>2a$, we first note that $\left|r-a\right|>0$, and thus in the whole integration range over $r'$ in Eq. (\ref{pnrecg02}) $r'>r-a>a$. Therefore, $\left|r'-a\right|>0$ as well. Hence, Eq. (\ref{pnrecg02}) may be written as
\begin{align}
p_{n+2}(r)=\frac{1}{4a^{2}}\int^{r+a}_{r-a}dr'\int^{r'+a}_{r'-a}dr''\frac{r''}{r}p_{n}\left(r''\right)dr'' .
\end{align}
Changing the order of integration yields
\begin{align}
p_{n+2}(r)=\frac{1}{4a^{2}}\left[\int^{r}_{r-2a}dr''\int^{r''+a}_{r-a}dr'+\int^{r+2a}_{r}dr''\int^{r+a}_{r''-a}dr'\right]\frac{r''}{r}p_{n}\left(r''\right)dr'' .
\end{align}
Performing the integrals over $r'$ yields
\begin{align}
p_{n+2}(r)=\frac{1}{4a^{2}}\left[\int^{r}_{r-2a}dr''\left(r''-r+2a\right)+\int^{r+2a}_{r}dr''\left(r-r''+2a\right)\right]\frac{r''}{r}p_{n}\left(r''\right)dr'' .
\end{align}
The derivative with respect to $r$ is
\begin{align}
\frac{\partial p_{n+2}(r)}{\partial r}=-\frac{1}{4a^{2}r^{2}}\left[\int^{r}_{r-2a}dr''\left(r''+2a\right)+\int^{r+2a}_{r}dr''\left(-r''+2a\right)\right]r''p_{n}\left(r''\right)dr'' .
\end{align}
In the second integral we change the integration variable to $r''-2a$ such that
\begin{align}
\frac{\partial p_{n+2}(r)}{\partial r}=-\frac{1}{4a^{2}r^{2}}\int^{r}_{r-2a}dr''\left(r''+2a\right)r''\left[p_{n}(r'')-p_{n}(r''+2a)\right] .
\end{align}
Since $p_{n}(r)$ is a non-increasing function of $r$ by the induction assumption, the integrand is non-negative, and thus the derivative of $p_{n+2}(r)$ is non-positive as required.

\subsection{$2a>r>a$}
In this case, we find that Eq. (\ref{pnrecg02}) reads
\begin{align}
p_{n+2}(r)=\frac{1}{4a^{2}}\left[\int^{r+a}_{a}dr'\int^{r'+a}_{r'-a}dr''+\int^{a}_{r-a}dr'\int^{r'+a}_{a-r}dr''\right]\frac{r''}{r}p_{n}\left(r''\right) .
\end{align}
Changing the order of integration yields
\begin{align}
&p_{n+2}(r)=\nonumber\\
&\frac{1}{4a^{2}}\left[\int^{2a-r}_{0}dr''\int^{r''+a}_{a-r''}dr'+\int^{r}_{2a-r}dr''\int^{r''+a}_{r-a}dr'+\int^{r+2a}_{r}dr''\int^{r+a}_{r''-a}dr'\right]\frac{r''}{r}p_{n}\left(r''\right) .
\end{align}
Performing the integrals over $r'$ yields
\begin{align}
&p_{n+2}(r)=\nonumber\\
&\frac{1}{4a^{2}}\left[\int^{2a-r}_{0}dr''2r''+\int^{r}_{2a-r}dr''\left(r''-r+2a\right)+\int^{r+2a}_{r}dr''\left(r-r''+2a\right)\right]\frac{r''}{r}p_{n}\left(r''\right) .
\end{align}
The derivative with respect to $r$ is
\begin{align}
&\frac{\partial p_{n+2}(r)}{\partial r}=\nonumber\\
&-\frac{1}{4a^{2}r^{2}}
\left[\int^{2a-r}_{0}dr''2r''+\int^{r}_{2a-r}dr''\left(r''+2a\right)+\int^{r+2a}_{r}dr''\left(-r''+2a\right)\right]r''p_{n}\left(r''\right) .
\end{align}
In the third integral we change the integration variable to $r''-2a$ such that
\begin{align}
&\frac{\partial p_{n+2}(r)}{\partial r}=\nonumber\\
&-\frac{1}{4a^{2}r^{2}}
\left\{2\int^{2a-r}_{0}dr''r''^{2}p_{n}(r'')+\int^{r}_{2a-r}dr''\left(r''+2a\right)r''\left[p_{n}\left(r''\right)-p_{n}\left(r''+2a\right)\right]\right\} .
\end{align}
As before, the integrand is positive, and thus $p_{n+2}(r)$ is a non-increasing function of $r$.

\subsection{$r<a$} 
In this last case we find that Eq. (\ref{pnrecg02}) reads
\begin{align}
p_{n+2}(r)=\frac{1}{4a^{2}}\left[\int^{r+a}_{a}dr'\int^{r'+a}_{r'-a}dr''+\int^{a}_{a-r}dr'\int^{r'+a}_{a-r'}dr''\right]\frac{r''}{r}p_{n}\left(r''\right) .
\end{align}
Changing the order of integration yields
\begin{align}
p_{n+2}(r)=\frac{1}{4a^{2}}\left[\int^{r}_{0}dr''\int^{a+r''}_{a-r''}dr'+\int^{2a-r}_{r}dr''\int^{r+a}_{a-r}dr'+\int^{r+2a}_{2a-r}dr''\int^{r+a}_{r''-a}dr'\right]\frac{r''}{r}p_{n}\left(r''\right) .
\end{align}
Integrating over $r'$ yields
\begin{align}
p_{n+2}(r)=\frac{1}{4a^{2}}\left[2\int^{r}_{0}dr''r''+2r\int^{2a-r}_{r}dr''+\int^{r+2a}_{2a-r}dr''\left(r-r''+2a\right)\right]\frac{r''}{r}p_{n}\left(r''\right) .
\end{align}
The derivative is
\begin{align}
&\frac{\partial p_{n+2}(r)}{\partial r}=-\frac{1}{4a^{2}r^{2}}\left[2\int^{r}_{0}dr''r''+\int^{r+2a}_{2a-r}dr''\left(-r''+2a\right)\right]r''p_{n}\left(r''\right)=\nonumber\\
&=-\frac{1}{4a^{2}r^{2}}\left[2\int^{r}_{0}dr''r''+\int^{r+2a}_{2a}dr''\left(-r''+2a\right)+\int^{2a}_{2a-r}dr''\left(-r''+2a\right)\right]r''p_{n}\left(r''\right) .
\end{align}
In the third integral we change the integration variable to $r''-2a$, and in the fourth integral we change it to $2a-r''$ such that
\begin{align}
&\frac{\partial p_{n+2}(r)}{\partial r}=\nonumber\\
&=-\frac{1}{4a^{2}r^{2}}\int^{r}_{0}dr''\left[2r''^{2}p_{n}(r)-r''\left(r''+2a\right)p_{n}\left(r''+2a\right)+r''\left(2a-r''\right)p_{n}\left(2a-r''\right)\right]=\nonumber\\
&=-\frac{1}{4a^{2}r^{2}}\int^{r}_{0}dr''\left\{r''^{2}\left[p_{n}(r)-p_{n}\left(r''+2a\right)+p_{n}(r)-p_{n}\left(2a-r''\right)\right]+\right.\nonumber\\
&\left.+2ar''\left[p_{n}\left(2a-r''\right)-p_{n}\left(2a+r''\right)\right]\right\} .
\end{align}
Since $r''<r<a$ we find that $2a-r''>r$ and thus the integrand is positive, and as before the derivative is non-positive.

In conclusion, we proved that if $\gamma_{f}=0$, i.e. there are no positive directional correlations, there can be at most one peak in the displacement distribution. Therefore, if there is more than one peak in the displacement distribution, this implies the existence of positive directional correlations.

\section{Summary}
\label{sec_summary}
In this paper we analyzed the motion of a single random walker with a specific type of persistence and external Gaussian noise. At each step it either moves in the exact same fashion (distance and direction) as in its previous move, makes the exact opposite move, or moves in a random direction. The timing of the different steps could be correlated. We derived an exact expression for the PDF of the walker's location for general time-correlations and general step-length distribution.

We explicitly evaluated the PDF for three types of step-length distributions and three types of time-correlations. We found that under certain conditions the PDF exhibits peaks at specific displacements, and that these peaks are more pronounced for narrower step-length distributions and higher persistence, i.e. propensity to continue in the same direction as before. The heights of these peaks decrease exponentially with the displacement if there are no correlations between the timing of the steps, but if correlations exist the height of the peaks could even increase with the displacement. 

Furthermore, we showed analytically that a necessary condition for the appearance of these peaks is a positive persistence, regardless of the time-correlations. Although our model is rather simplistic, we believe that this conclusion can be generalized. Namely, if the PDF exhibits peaks at various values of the displacement then the particle is persistent, i.e. it has a positive velocity autocorrelation.

One way to expand this work would be to consider a more realistic type of persistence, such as an Orenstein-Uhlenbeck process or Levy walks. Investigating more thoroughly the effect of the temporal correlations on the relative height of the peaks would also be an interesting avenue of research.

Our results may be used to understand the microscopic processes underlying the overall observed random walk. At long times many of the details of the random motion are averaged out, but they are present in the short time behaviour. Using our results, and expanding the simple model we present here, one may look for non-monotonicity; or, if such non-monotonicity is not observed at short times, this implies a bound on the orientational correlations. By looking for peaks in the displacement distribution in experiments, one may obtain data about the directional correlations of the particles. Moreover, the heights and the separation of the peaks give further insight into the underlying microscopic processes.

\section*{Acknowledgments}
We thank Niv Gov, Ralf Meltzer and Michael Urbakh for helpful discussions. This research was supported by the US-Israel Binational Science Foundation, by the Israel Science Foundation Grant No. 968/16, and by the joint Tel Aviv University - Potsdam University scholarship.

\bibliography{references}

\begin{thebibliography}{69}
\expandafter\ifx\csname natexlab\endcsname\relax\def\natexlab#1{#1}\fi
\expandafter\ifx\csname bibnamefont\endcsname\relax
  \def\bibnamefont#1{#1}\fi
\expandafter\ifx\csname bibfnamefont\endcsname\relax
  \def\bibfnamefont#1{#1}\fi
\expandafter\ifx\csname citenamefont\endcsname\relax
  \def\citenamefont#1{#1}\fi
\expandafter\ifx\csname url\endcsname\relax
  \def\url#1{\texttt{#1}}\fi
\expandafter\ifx\csname urlprefix\endcsname\relax\def\urlprefix{URL }\fi
\providecommand{\bibinfo}[2]{#2}
\providecommand{\eprint}[2][]{\url{#2}}

\bibitem[{\citenamefont{de~Groot and Grubm{\" u}ller}(2005)}]{Groot2005}
\bibinfo{author}{\bibfnamefont{B.}~\bibnamefont{de~Groot}} \bibnamefont{and}
  \bibinfo{author}{\bibfnamefont{L.}~\bibnamefont{Grubm{\" u}ller}},
  \bibinfo{journal}{Curr. Opin. Struct. Biol.} \textbf{\bibinfo{volume}{15}},
  \bibinfo{pages}{176} (\bibinfo{year}{2005}).

\bibitem[{\citenamefont{H{\" o}fling and Franosch}(2014)}]{Hofling2014}
\bibinfo{author}{\bibfnamefont{F.}~\bibnamefont{H{\" o}fling}}
  \bibnamefont{and} \bibinfo{author}{\bibfnamefont{T.}~\bibnamefont{Franosch}},
  \bibinfo{journal}{Rep. Prog. Phys.} \textbf{\bibinfo{volume}{76}},
  \bibinfo{pages}{046602} (\bibinfo{year}{2014}).

\bibitem[{\citenamefont{Bechinger et~al.}(2016)\citenamefont{Bechinger,
  Di~Leonardo, L{\" o}wen, Reichhardt, Volpe, and Volpe}}]{Bechinger2016}
\bibinfo{author}{\bibfnamefont{C.}~\bibnamefont{Bechinger}},
  \bibinfo{author}{\bibfnamefont{R.}~\bibnamefont{Di~Leonardo}},
  \bibinfo{author}{\bibfnamefont{H.}~\bibnamefont{L{\" o}wen}},
  \bibinfo{author}{\bibfnamefont{C.}~\bibnamefont{Reichhardt}},
  \bibinfo{author}{\bibfnamefont{G.}~\bibnamefont{Volpe}}, \bibnamefont{and}
  \bibinfo{author}{\bibfnamefont{G.}~\bibnamefont{Volpe}},
  \bibinfo{journal}{Rev. Mod. Phys.} \textbf{\bibinfo{volume}{88}},
  \bibinfo{pages}{045006} (\bibinfo{year}{2016}).

\bibitem[{\citenamefont{Metzler et~al.}(2016)\citenamefont{Metzler, Jeon, and
  Cherstvy}}]{Metzler2016}
\bibinfo{author}{\bibfnamefont{R.}~\bibnamefont{Metzler}},
  \bibinfo{author}{\bibfnamefont{J.}~\bibnamefont{Jeon}}, \bibnamefont{and}
  \bibinfo{author}{\bibfnamefont{A.}~\bibnamefont{Cherstvy}},
  \bibinfo{journal}{BBA-Rev. Biomembranes} \textbf{\bibinfo{volume}{1858}},
  \bibinfo{pages}{2451} (\bibinfo{year}{2016}).

\bibitem[{\citenamefont{Hakim and Silberzan}(2017)}]{Hakim2017}
\bibinfo{author}{\bibfnamefont{V.}~\bibnamefont{Hakim}} \bibnamefont{and}
  \bibinfo{author}{\bibfnamefont{P.}~\bibnamefont{Silberzan}},
  \bibinfo{journal}{Rep. Prog. Phys.} \textbf{\bibinfo{volume}{80}},
  \bibinfo{pages}{076601} (\bibinfo{year}{2017}).

\bibitem[{\citenamefont{Norregaard et~al.}(2017)\citenamefont{Norregaard,
  Metzler, Ritter, Berg-S{\o}rensen, and Oddershede}}]{Norregaard2017}
\bibinfo{author}{\bibfnamefont{K.}~\bibnamefont{Norregaard}},
  \bibinfo{author}{\bibfnamefont{R.}~\bibnamefont{Metzler}},
  \bibinfo{author}{\bibfnamefont{C.}~\bibnamefont{Ritter}},
  \bibinfo{author}{\bibfnamefont{K.}~\bibnamefont{Berg-S{\o}rensen}},
  \bibnamefont{and}
  \bibinfo{author}{\bibfnamefont{L.}~\bibnamefont{Oddershede}},
  \bibinfo{journal}{Chem. Rev.} \textbf{\bibinfo{volume}{117}},
  \bibinfo{pages}{4342} (\bibinfo{year}{2017}).

\bibitem[{\citenamefont{Hasnain and Bandyopadhyay}(2015)}]{Hasnain2015}
\bibinfo{author}{\bibfnamefont{S.}~\bibnamefont{Hasnain}} \bibnamefont{and}
  \bibinfo{author}{\bibfnamefont{P.}~\bibnamefont{Bandyopadhyay}},
  \bibinfo{journal}{J. Chem. Phys.} \textbf{\bibinfo{volume}{143}},
  \bibinfo{pages}{114104} (\bibinfo{year}{2015}).

\bibitem[{\citenamefont{Detcheverry}(2015)}]{Detcheverry2015}
\bibinfo{author}{\bibfnamefont{F.}~\bibnamefont{Detcheverry}},
  \bibinfo{journal}{Europhys. Lett.} \textbf{\bibinfo{volume}{111}},
  \bibinfo{pages}{60002} (\bibinfo{year}{2015}).

\bibitem[{\citenamefont{Rupprecht et~al.}(2016)\citenamefont{Rupprecht,
  B{\'e}nichou, and Voituriez}}]{Rupprecht2016}
\bibinfo{author}{\bibfnamefont{J.}~\bibnamefont{Rupprecht}},
  \bibinfo{author}{\bibfnamefont{O.}~\bibnamefont{B{\'e}nichou}},
  \bibnamefont{and}
  \bibinfo{author}{\bibfnamefont{R.}~\bibnamefont{Voituriez}},
  \bibinfo{journal}{Phys. Rev. E} \textbf{\bibinfo{volume}{94}},
  \bibinfo{pages}{012117} (\bibinfo{year}{2016}).

\bibitem[{\citenamefont{Detcheverry}(2017)}]{Detcheverry2017}
\bibinfo{author}{\bibfnamefont{F.}~\bibnamefont{Detcheverry}},
  \bibinfo{journal}{Phys. Rev. E} \textbf{\bibinfo{volume}{96}},
  \bibinfo{pages}{012415} (\bibinfo{year}{2017}).

\bibitem[{\citenamefont{Sevilla}(2016)}]{Sevilla2016}
\bibinfo{author}{\bibfnamefont{F.}~\bibnamefont{Sevilla}},
  \bibinfo{journal}{Phys. Rev. E} \textbf{\bibinfo{volume}{94}},
  \bibinfo{pages}{062120} (\bibinfo{year}{2016}).

\bibitem[{\citenamefont{Ariel et~al.}(2017)\citenamefont{Ariel, Be'er, and
  Reynolds}}]{Ariel2017}
\bibinfo{author}{\bibfnamefont{G.}~\bibnamefont{Ariel}},
  \bibinfo{author}{\bibfnamefont{A.}~\bibnamefont{Be'er}}, \bibnamefont{and}
  \bibinfo{author}{\bibfnamefont{A.}~\bibnamefont{Reynolds}},
  \bibinfo{journal}{Phys. Rev. Lett.} \textbf{\bibinfo{volume}{118}},
  \bibinfo{pages}{228102} (\bibinfo{year}{2017}).

\bibitem[{\citenamefont{Fedotov and Korabel}(2017)}]{Fedotov2017}
\bibinfo{author}{\bibfnamefont{S.}~\bibnamefont{Fedotov}} \bibnamefont{and}
  \bibinfo{author}{\bibfnamefont{N.}~\bibnamefont{Korabel}},
  \bibinfo{journal}{Phys. Rev. E} \textbf{\bibinfo{volume}{95}},
  \bibinfo{pages}{030107} (\bibinfo{year}{2017}).

\bibitem[{\citenamefont{Wioland et~al.}(2016)\citenamefont{Wioland, Lushi, and
  Goldstein}}]{Wioland2016}
\bibinfo{author}{\bibfnamefont{H.}~\bibnamefont{Wioland}},
  \bibinfo{author}{\bibfnamefont{E.}~\bibnamefont{Lushi}}, \bibnamefont{and}
  \bibinfo{author}{\bibfnamefont{R.}~\bibnamefont{Goldstein}},
  \bibinfo{journal}{New J. Phys.} \textbf{\bibinfo{volume}{18}},
  \bibinfo{pages}{075002} (\bibinfo{year}{2016}).

\bibitem[{\citenamefont{Stenhammer et~al.}(2017)\citenamefont{Stenhammer,
  Nardini, Nash, Marenduzzo, and Mozorov}}]{Stenhammer2017}
\bibinfo{author}{\bibfnamefont{J.}~\bibnamefont{Stenhammer}},
  \bibinfo{author}{\bibfnamefont{C.}~\bibnamefont{Nardini}},
  \bibinfo{author}{\bibfnamefont{R.}~\bibnamefont{Nash}},
  \bibinfo{author}{\bibfnamefont{D.}~\bibnamefont{Marenduzzo}},
  \bibnamefont{and} \bibinfo{author}{\bibfnamefont{A.}~\bibnamefont{Mozorov}},
  \bibinfo{journal}{Phys. Rev. Lett.} \textbf{\bibinfo{volume}{119}},
  \bibinfo{pages}{028005} (\bibinfo{year}{2017}).

\bibitem[{\citenamefont{Berthier and Kurchan}(2013)}]{Berthier2013}
\bibinfo{author}{\bibfnamefont{L.}~\bibnamefont{Berthier}} \bibnamefont{and}
  \bibinfo{author}{\bibfnamefont{J.}~\bibnamefont{Kurchan}},
  \bibinfo{journal}{Nat. Phys.} \textbf{\bibinfo{volume}{9}},
  \bibinfo{pages}{310} (\bibinfo{year}{2013}).

\bibitem[{\citenamefont{Viscek et~al.}(1995)\citenamefont{Viscek, Czir{\'o}k,
  Ben-Jacob, Cohen, and Shochet}}]{Viscek1995}
\bibinfo{author}{\bibfnamefont{T.}~\bibnamefont{Viscek}},
  \bibinfo{author}{\bibfnamefont{A.}~\bibnamefont{Czir{\'o}k}},
  \bibinfo{author}{\bibfnamefont{E.}~\bibnamefont{Ben-Jacob}},
  \bibinfo{author}{\bibfnamefont{I.}~\bibnamefont{Cohen}}, \bibnamefont{and}
  \bibinfo{author}{\bibfnamefont{O.}~\bibnamefont{Shochet}},
  \bibinfo{journal}{Phys. Rev. Lett.} \textbf{\bibinfo{volume}{75}},
  \bibinfo{pages}{1226} (\bibinfo{year}{1995}).

\bibitem[{\citenamefont{Sep{\'u}lveda et~al.}(2013)\citenamefont{Sep{\'u}lveda,
  Petitjean, Cochet, Grasland-Mongrain, Silberzan, and Hakim}}]{Sepulveda2013}
\bibinfo{author}{\bibfnamefont{N.}~\bibnamefont{Sep{\'u}lveda}},
  \bibinfo{author}{\bibfnamefont{L.}~\bibnamefont{Petitjean}},
  \bibinfo{author}{\bibfnamefont{O.}~\bibnamefont{Cochet}},
  \bibinfo{author}{\bibfnamefont{E.}~\bibnamefont{Grasland-Mongrain}},
  \bibinfo{author}{\bibfnamefont{P.}~\bibnamefont{Silberzan}},
  \bibnamefont{and} \bibinfo{author}{\bibfnamefont{V.}~\bibnamefont{Hakim}},
  \bibinfo{journal}{PLoS Comput. Biol.} \textbf{\bibinfo{volume}{9}},
  \bibinfo{pages}{1002944} (\bibinfo{year}{2013}).

\bibitem[{\citenamefont{Gro{\ss}mann et~al.}(2016)\citenamefont{Gro{\ss}mann,
  Peruani, and B{\"a}r}}]{Grossmann2016}
\bibinfo{author}{\bibfnamefont{R.}~\bibnamefont{Gro{\ss}mann}},
  \bibinfo{author}{\bibfnamefont{F.}~\bibnamefont{Peruani}}, \bibnamefont{and}
  \bibinfo{author}{\bibfnamefont{M.}~\bibnamefont{B{\"a}r}},
  \bibinfo{journal}{Phys. Rev. E} \textbf{\bibinfo{volume}{94}},
  \bibinfo{pages}{050602} (\bibinfo{year}{2016}).

\bibitem[{\citenamefont{Liebchen and Levis}(2017)}]{Liebchen2017}
\bibinfo{author}{\bibfnamefont{B.}~\bibnamefont{Liebchen}} \bibnamefont{and}
  \bibinfo{author}{\bibfnamefont{D.}~\bibnamefont{Levis}},
  \bibinfo{journal}{Phys. Rev. Lett,} \textbf{\bibinfo{volume}{119}},
  \bibinfo{pages}{058002} (\bibinfo{year}{2017}).

\bibitem[{\citenamefont{Zimmerman et~al.}(2016)\citenamefont{Zimmerman, Camley,
  Rappel, and Levine}}]{Zimmermann2016}
\bibinfo{author}{\bibfnamefont{J.}~\bibnamefont{Zimmerman}},
  \bibinfo{author}{\bibfnamefont{B.}~\bibnamefont{Camley}},
  \bibinfo{author}{\bibfnamefont{W.}~\bibnamefont{Rappel}}, \bibnamefont{and}
  \bibinfo{author}{\bibfnamefont{H.}~\bibnamefont{Levine}},
  \bibinfo{journal}{Proc. Natl. Acad. Sci. USA} \textbf{\bibinfo{volume}{113}},
  \bibinfo{pages}{2660} (\bibinfo{year}{2016}).

\bibitem[{\citenamefont{Farhadifar et~al.}(2007)\citenamefont{Farhadifar,
  R{\"o}per, Aigouy, Eaton, and J{\"u}licher}}]{Farhadifar2007}
\bibinfo{author}{\bibfnamefont{R.}~\bibnamefont{Farhadifar}},
  \bibinfo{author}{\bibfnamefont{J.}~\bibnamefont{R{\"o}per}},
  \bibinfo{author}{\bibfnamefont{B.}~\bibnamefont{Aigouy}},
  \bibinfo{author}{\bibfnamefont{S.}~\bibnamefont{Eaton}}, \bibnamefont{and}
  \bibinfo{author}{\bibfnamefont{F.}~\bibnamefont{J{\"u}licher}},
  \bibinfo{journal}{Curr. Biol.} \textbf{\bibinfo{volume}{17}},
  \bibinfo{pages}{2095} (\bibinfo{year}{2007}).

\bibitem[{\citenamefont{Staple et~al.}(2010)\citenamefont{Staple, Farhadifar,
  R{\"o}per, Aigouy, Eaton, and J{\"u}licher}}]{Staple2010}
\bibinfo{author}{\bibfnamefont{D.}~\bibnamefont{Staple}},
  \bibinfo{author}{\bibfnamefont{R.}~\bibnamefont{Farhadifar}},
  \bibinfo{author}{\bibfnamefont{J.}~\bibnamefont{R{\"o}per}},
  \bibinfo{author}{\bibfnamefont{B.}~\bibnamefont{Aigouy}},
  \bibinfo{author}{\bibfnamefont{S.}~\bibnamefont{Eaton}}, \bibnamefont{and}
  \bibinfo{author}{\bibfnamefont{F.}~\bibnamefont{J{\"u}licher}},
  \bibinfo{journal}{Eur. Phys. J. E} \textbf{\bibinfo{volume}{33}},
  \bibinfo{pages}{117} (\bibinfo{year}{2010}).

\bibitem[{\citenamefont{S{\'a}ndor et~al.}(2017)\citenamefont{S{\'a}ndor,
  Lib{\'a}l, Reichhardt, and Reichhardt}}]{Sandor2017}
\bibinfo{author}{\bibfnamefont{C.}~\bibnamefont{S{\'a}ndor}},
  \bibinfo{author}{\bibfnamefont{A.}~\bibnamefont{Lib{\'a}l}},
  \bibinfo{author}{\bibfnamefont{C.}~\bibnamefont{Reichhardt}},
  \bibnamefont{and}
  \bibinfo{author}{\bibfnamefont{C.}~\bibnamefont{Reichhardt}},
  \bibinfo{journal}{Phys. Rev. E} \textbf{\bibinfo{volume}{95}},
  \bibinfo{pages}{032606} (\bibinfo{year}{2017}).

\bibitem[{\citenamefont{Reichhardt and
  Reichhardt}(2014{\natexlab{a}})}]{Reichhardt2014}
\bibinfo{author}{\bibfnamefont{C.}~\bibnamefont{Reichhardt}} \bibnamefont{and}
  \bibinfo{author}{\bibfnamefont{C.}~\bibnamefont{Reichhardt}},
  \bibinfo{journal}{Soft Matter} \textbf{\bibinfo{volume}{10}},
  \bibinfo{pages}{7502} (\bibinfo{year}{2014}{\natexlab{a}}).

\bibitem[{\citenamefont{Zacherson et~al.}(2017)\citenamefont{Zacherson, Wolff,
  Whitchurch, and Toth}}]{Zacherson2017}
\bibinfo{author}{\bibfnamefont{C.}~\bibnamefont{Zacherson}},
  \bibinfo{author}{\bibfnamefont{C.}~\bibnamefont{Wolff}},
  \bibinfo{author}{\bibfnamefont{C.}~\bibnamefont{Whitchurch}},
  \bibnamefont{and} \bibinfo{author}{\bibfnamefont{M.}~\bibnamefont{Toth}},
  \bibinfo{journal}{Phys. Rev. E} \textbf{\bibinfo{volume}{95}},
  \bibinfo{pages}{012408} (\bibinfo{year}{2017}).

\bibitem[{\citenamefont{Reichhardt and
  Reichhardt}(2014{\natexlab{b}})}]{Reichhardt2014b}
\bibinfo{author}{\bibfnamefont{C.}~\bibnamefont{Reichhardt}} \bibnamefont{and}
  \bibinfo{author}{\bibfnamefont{C.}~\bibnamefont{Reichhardt}},
  \bibinfo{journal}{Phys. Rev. E} \textbf{\bibinfo{volume}{90}},
  \bibinfo{pages}{012701} (\bibinfo{year}{2014}{\natexlab{b}}).

\bibitem[{\citenamefont{Lam et~al.}(2015)\citenamefont{Lam, Schindler, and
  Dauchot}}]{Lam2015}
\bibinfo{author}{\bibfnamefont{K.}~\bibnamefont{Lam}},
  \bibinfo{author}{\bibfnamefont{M.}~\bibnamefont{Schindler}},
  \bibnamefont{and} \bibinfo{author}{\bibfnamefont{O.}~\bibnamefont{Dauchot}},
  \bibinfo{journal}{New J. Phys.} \textbf{\bibinfo{volume}{17}},
  \bibinfo{pages}{113056} (\bibinfo{year}{2015}).

\bibitem[{\citenamefont{Graf and Frey}(2017)}]{Graf2017}
\bibinfo{author}{\bibfnamefont{I.}~\bibnamefont{Graf}} \bibnamefont{and}
  \bibinfo{author}{\bibfnamefont{E.}~\bibnamefont{Frey}},
  \bibinfo{journal}{Phys. Rev. Lett.} \textbf{\bibinfo{volume}{188}},
  \bibinfo{pages}{128101} (\bibinfo{year}{2017}).

\bibitem[{\citenamefont{Illien~P and R}(2015)}]{Illien2015}
\bibinfo{author}{\bibfnamefont{O.~G.} \bibnamefont{Illien~P},
  \bibfnamefont{B{\'e}nichou~O}} \bibnamefont{and}
  \bibinfo{author}{\bibfnamefont{V.}~\bibnamefont{R}}, \bibinfo{journal}{J.
  Stat. Mech.} \textbf{\bibinfo{volume}{2015}}, \bibinfo{pages}{P11016}
  (\bibinfo{year}{2015}).

\bibitem[{\citenamefont{Berg and Turner}(1990)}]{Berg1990}
\bibinfo{author}{\bibfnamefont{H.}~\bibnamefont{Berg}} \bibnamefont{and}
  \bibinfo{author}{\bibfnamefont{L.}~\bibnamefont{Turner}},
  \bibinfo{journal}{Biophys. J.} \textbf{\bibinfo{volume}{58}},
  \bibinfo{pages}{919} (\bibinfo{year}{1990}).

\bibitem[{\citenamefont{Ghosh et~al.}(2015{\natexlab{a}})\citenamefont{Ghosh,
  Cherstvy, and Metzler}}]{Ghosh2015}
\bibinfo{author}{\bibfnamefont{S.}~\bibnamefont{Ghosh}},
  \bibinfo{author}{\bibfnamefont{A.}~\bibnamefont{Cherstvy}}, \bibnamefont{and}
  \bibinfo{author}{\bibfnamefont{R.}~\bibnamefont{Metzler}},
  \bibinfo{journal}{Phys. Chem. Chem. Phys.} \textbf{\bibinfo{volume}{17}},
  \bibinfo{pages}{1847} (\bibinfo{year}{2015}{\natexlab{a}}).

\bibitem[{\citenamefont{Schulz et~al.}(2011)\citenamefont{Schulz, Kolomeisky,
  and Frey}}]{Schulz2011}
\bibinfo{author}{\bibfnamefont{J.}~\bibnamefont{Schulz}},
  \bibinfo{author}{\bibfnamefont{A.}~\bibnamefont{Kolomeisky}},
  \bibnamefont{and} \bibinfo{author}{\bibfnamefont{E.}~\bibnamefont{Frey}},
  \bibinfo{journal}{Europhys. Lett.} \textbf{\bibinfo{volume}{95}},
  \bibinfo{pages}{30004} (\bibinfo{year}{2011}).

\bibitem[{\citenamefont{Hermann et~al.}(2017)\citenamefont{Hermann, Metzler,
  and Engbert}}]{Hermann2017}
\bibinfo{author}{\bibfnamefont{C.}~\bibnamefont{Hermann}},
  \bibinfo{author}{\bibfnamefont{R.}~\bibnamefont{Metzler}}, \bibnamefont{and}
  \bibinfo{author}{\bibfnamefont{R.}~\bibnamefont{Engbert}},
  \bibinfo{journal}{Sci. Rep.} \textbf{\bibinfo{volume}{7}},
  \bibinfo{pages}{12958} (\bibinfo{year}{2017}).

\bibitem[{\citenamefont{Tchen}(1952)}]{Tchen1952}
\bibinfo{author}{\bibfnamefont{C.}~\bibnamefont{Tchen}}, \bibinfo{journal}{J.
  Chem. Phys.} \textbf{\bibinfo{volume}{20}}, \bibinfo{pages}{214}
  (\bibinfo{year}{1952}).

\bibitem[{\citenamefont{Kareiva and Shigesada}(1983)}]{Kareiva1983}
\bibinfo{author}{\bibfnamefont{P.}~\bibnamefont{Kareiva}} \bibnamefont{and}
  \bibinfo{author}{\bibfnamefont{N.}~\bibnamefont{Shigesada}},
  \bibinfo{journal}{Oeocologia} \textbf{\bibinfo{volume}{56}},
  \bibinfo{pages}{234} (\bibinfo{year}{1983}).

\bibitem[{\citenamefont{Bogu{\~n}a et~al.}(1998)\citenamefont{Bogu{\~n}a,
  Porr{\'a}, and Masoliver}}]{Boguna1998}
\bibinfo{author}{\bibfnamefont{M.}~\bibnamefont{Bogu{\~n}a}},
  \bibinfo{author}{\bibfnamefont{J.}~\bibnamefont{Porr{\'a}}},
  \bibnamefont{and}
  \bibinfo{author}{\bibfnamefont{J.}~\bibnamefont{Masoliver}},
  \bibinfo{journal}{Phys. Rev. E} \textbf{\bibinfo{volume}{58}},
  \bibinfo{pages}{6992} (\bibinfo{year}{1998}).

\bibitem[{\citenamefont{Romanczuk et~al.}(2012)\citenamefont{Romanczuk,
  B{\"a}r, Ebeling, Lindner, and Schminasky-Geier}}]{Romanczuk2012}
\bibinfo{author}{\bibfnamefont{P.}~\bibnamefont{Romanczuk}},
  \bibinfo{author}{\bibfnamefont{M.}~\bibnamefont{B{\"a}r}},
  \bibinfo{author}{\bibfnamefont{W.}~\bibnamefont{Ebeling}},
  \bibinfo{author}{\bibfnamefont{B.}~\bibnamefont{Lindner}}, \bibnamefont{and}
  \bibinfo{author}{\bibfnamefont{L.}~\bibnamefont{Schminasky-Geier}},
  \bibinfo{journal}{Eur. Phys. J.} \textbf{\bibinfo{volume}{202}},
  \bibinfo{pages}{1} (\bibinfo{year}{2012}).

\bibitem[{\citenamefont{Ghosh et~al.}(2015{\natexlab{b}})\citenamefont{Ghosh,
  Li, Marchgiani, and Marchesoni}}]{Ghosh2015b}
\bibinfo{author}{\bibfnamefont{P.}~\bibnamefont{Ghosh}},
  \bibinfo{author}{\bibfnamefont{Y.}~\bibnamefont{Li}},
  \bibinfo{author}{\bibfnamefont{G.}~\bibnamefont{Marchgiani}},
  \bibnamefont{and}
  \bibinfo{author}{\bibfnamefont{F.}~\bibnamefont{Marchesoni}},
  \bibinfo{journal}{J. Chem. Phys.} \textbf{\bibinfo{volume}{143}},
  \bibinfo{pages}{211101} (\bibinfo{year}{2015}{\natexlab{b}}).

\bibitem[{\citenamefont{Tahir-Kheli and Elliot}(1983)}]{Tahir1983}
\bibinfo{author}{\bibfnamefont{R.}~\bibnamefont{Tahir-Kheli}} \bibnamefont{and}
  \bibinfo{author}{\bibfnamefont{R.}~\bibnamefont{Elliot}},
  \bibinfo{journal}{Phys. Rev. B} \textbf{\bibinfo{volume}{27}},
  \bibinfo{pages}{844} (\bibinfo{year}{1983}).

\bibitem[{\citenamefont{Tahir-Kheli}(1983)}]{Tahir1983b}
\bibinfo{author}{\bibfnamefont{R.}~\bibnamefont{Tahir-Kheli}},
  \bibinfo{journal}{Phys. Rev. B} \textbf{\bibinfo{volume}{27}},
  \bibinfo{pages}{7229} (\bibinfo{year}{1983}).

\bibitem[{\citenamefont{Lettinga and Grelet}(2007)}]{Lettinga2007}
\bibinfo{author}{\bibfnamefont{M.}~\bibnamefont{Lettinga}} \bibnamefont{and}
  \bibinfo{author}{\bibfnamefont{E.}~\bibnamefont{Grelet}},
  \bibinfo{journal}{Phys. Rev. Lett.} \textbf{\bibinfo{volume}{99}},
  \bibinfo{pages}{197802} (\bibinfo{year}{2007}).

\bibitem[{\citenamefont{Pouget et~al.}(2011)\citenamefont{Pouget, Grelet, and
  Lettinga}}]{Pouget2011}
\bibinfo{author}{\bibfnamefont{E.}~\bibnamefont{Pouget}},
  \bibinfo{author}{\bibfnamefont{E.}~\bibnamefont{Grelet}}, \bibnamefont{and}
  \bibinfo{author}{\bibfnamefont{M.}~\bibnamefont{Lettinga}},
  \bibinfo{journal}{Phys. Rev. E} \textbf{\bibinfo{volume}{84}},
  \bibinfo{pages}{041704} (\bibinfo{year}{2011}).

\bibitem[{\citenamefont{Naderi et~al.}(2013)\citenamefont{Naderi, Pouget,
  Ballesta, van~der Schoot, Lettinga, and Grelet}}]{Naderi2013}
\bibinfo{author}{\bibfnamefont{S.}~\bibnamefont{Naderi}},
  \bibinfo{author}{\bibfnamefont{E.}~\bibnamefont{Pouget}},
  \bibinfo{author}{\bibfnamefont{P.}~\bibnamefont{Ballesta}},
  \bibinfo{author}{\bibfnamefont{P.}~\bibnamefont{van~der Schoot}},
  \bibinfo{author}{\bibfnamefont{M.}~\bibnamefont{Lettinga}}, \bibnamefont{and}
  \bibinfo{author}{\bibfnamefont{E.}~\bibnamefont{Grelet}},
  \bibinfo{journal}{Phys. Rev. Lett.} \textbf{\bibinfo{volume}{111}},
  \bibinfo{pages}{037801} (\bibinfo{year}{2013}).

\bibitem[{\citenamefont{Sonn-Segev et~al.}(2017)\citenamefont{Sonn-Segev,
  Bernheim-Groswasser, and Roichman}}]{SonnSegev2017}
\bibinfo{author}{\bibfnamefont{A.}~\bibnamefont{Sonn-Segev}},
  \bibinfo{author}{\bibfnamefont{A.}~\bibnamefont{Bernheim-Groswasser}},
  \bibnamefont{and} \bibinfo{author}{\bibfnamefont{Y.}~\bibnamefont{Roichman}},
  \bibinfo{journal}{Soft Matter} \textbf{\bibinfo{volume}{13}},
  \bibinfo{pages}{7352} (\bibinfo{year}{2017}).

\bibitem[{\citenamefont{Goldstein}(1951)}]{Goldstein1951}
\bibinfo{author}{\bibfnamefont{S.}~\bibnamefont{Goldstein}},
  \bibinfo{journal}{Q. J. Mech. Appl. Math.} \textbf{\bibinfo{volume}{4}},
  \bibinfo{pages}{129} (\bibinfo{year}{1951}).

\bibitem[{\citenamefont{Uhlenbeck and Ornstein}(1930)}]{Uhlenbeck1930}
\bibinfo{author}{\bibfnamefont{G.}~\bibnamefont{Uhlenbeck}} \bibnamefont{and}
  \bibinfo{author}{\bibfnamefont{L.}~\bibnamefont{Ornstein}},
  \bibinfo{journal}{Phys. Rev. Lett.} \textbf{\bibinfo{volume}{36}},
  \bibinfo{pages}{823} (\bibinfo{year}{1930}).

\bibitem[{\citenamefont{Montroll and Weiss}(1965)}]{Montroll1965}
\bibinfo{author}{\bibfnamefont{E.}~\bibnamefont{Montroll}} \bibnamefont{and}
  \bibinfo{author}{\bibfnamefont{G.}~\bibnamefont{Weiss}}, \bibinfo{journal}{J.
  Math. Phys.} \textbf{\bibinfo{volume}{6}}, \bibinfo{pages}{167}
  (\bibinfo{year}{1965}).

\bibitem[{\citenamefont{Shlesinger et~al.}(1987)\citenamefont{Shlesinger, West,
  and Klafter}}]{Shlesinger1987}
\bibinfo{author}{\bibfnamefont{M.}~\bibnamefont{Shlesinger}},
  \bibinfo{author}{\bibfnamefont{B.}~\bibnamefont{West}}, \bibnamefont{and}
  \bibinfo{author}{\bibfnamefont{J.}~\bibnamefont{Klafter}},
  \bibinfo{journal}{Phys. Rev. Lett.} \textbf{\bibinfo{volume}{58}},
  \bibinfo{pages}{1100} (\bibinfo{year}{1987}).

\bibitem[{\citenamefont{Levy}(1939)}]{Levy1939}
\bibinfo{author}{\bibfnamefont{P.}~\bibnamefont{Levy}}, \bibinfo{journal}{B.
  Soc. Math. Fr.} \textbf{\bibinfo{volume}{67}}, \bibinfo{pages}{1}
  (\bibinfo{year}{1939}).

\bibitem[{\citenamefont{Rayleigh}(1880)}]{Rayleigh1880}
\bibinfo{author}{\bibfnamefont{F.}~\bibnamefont{Rayleigh}},
  \bibinfo{journal}{Philos. Mag.} \textbf{\bibinfo{volume}{10}},
  \bibinfo{pages}{73} (\bibinfo{year}{1880}).

\bibitem[{\citenamefont{de~Nigris et~al.}(2017)\citenamefont{de~Nigris,
  Carletti, and Lambiotte}}]{Nigris2017}
\bibinfo{author}{\bibfnamefont{S.}~\bibnamefont{de~Nigris}},
  \bibinfo{author}{\bibfnamefont{T.}~\bibnamefont{Carletti}}, \bibnamefont{and}
  \bibinfo{author}{\bibfnamefont{R.}~\bibnamefont{Lambiotte}},
  \bibinfo{journal}{Phys. Rev. E} \textbf{\bibinfo{volume}{95}},
  \bibinfo{pages}{022113} (\bibinfo{year}{2017}).

\bibitem[{\citenamefont{Codling et~al.}(2008)\citenamefont{Codling, Plank, and
  Benhamou}}]{Coding2008}
\bibinfo{author}{\bibfnamefont{E.}~\bibnamefont{Codling}},
  \bibinfo{author}{\bibfnamefont{M.}~\bibnamefont{Plank}}, \bibnamefont{and}
  \bibinfo{author}{\bibfnamefont{S.}~\bibnamefont{Benhamou}},
  \bibinfo{journal}{J. R. Soc. Interface} \textbf{\bibinfo{volume}{5}},
  \bibinfo{pages}{813} (\bibinfo{year}{2008}).

\bibitem[{\citenamefont{Metzler et~al.}(2014)\citenamefont{Metzler, Jeon,
  Cherstvy, and Barkai}}]{Metzler2014}
\bibinfo{author}{\bibfnamefont{R.}~\bibnamefont{Metzler}},
  \bibinfo{author}{\bibfnamefont{J.}~\bibnamefont{Jeon}},
  \bibinfo{author}{\bibfnamefont{A.}~\bibnamefont{Cherstvy}}, \bibnamefont{and}
  \bibinfo{author}{\bibfnamefont{E.}~\bibnamefont{Barkai}},
  \bibinfo{journal}{Phys. Chem. Chem. Phys.} \textbf{\bibinfo{volume}{16}},
  \bibinfo{pages}{24128} (\bibinfo{year}{2014}).

\bibitem[{\citenamefont{Sposini et~al.}(2018)\citenamefont{Sposini, Chechkin,
  Seno, Pagnini, and Metzler}}]{Sposini2018}
\bibinfo{author}{\bibfnamefont{V.}~\bibnamefont{Sposini}},
  \bibinfo{author}{\bibfnamefont{A.}~\bibnamefont{Chechkin}},
  \bibinfo{author}{\bibfnamefont{F.}~\bibnamefont{Seno}},
  \bibinfo{author}{\bibfnamefont{G.}~\bibnamefont{Pagnini}}, \bibnamefont{and}
  \bibinfo{author}{\bibfnamefont{R.}~\bibnamefont{Metzler}},
  \bibinfo{journal}{New J. Phys.} \textbf{\bibinfo{volume}{20}},
  \bibinfo{pages}{043044} (\bibinfo{year}{2018}).

\bibitem[{\citenamefont{Jakub and Metzler}(2018)}]{Jakub2018}
\bibinfo{author}{\bibfnamefont{S.}~\bibnamefont{Jakub}} \bibnamefont{and}
  \bibinfo{author}{\bibfnamefont{R.}~\bibnamefont{Metzler}},
  \bibinfo{journal}{New J. Phys.} \textbf{\bibinfo{volume}{20}},
  \bibinfo{pages}{023026} (\bibinfo{year}{2018}).

\bibitem[{\citenamefont{Gnacik et~al.}(2018)\citenamefont{Gnacik, Alsolami, and
  Burridge}}]{Gnacik2018}
\bibinfo{author}{\bibfnamefont{M.}~\bibnamefont{Gnacik}},
  \bibinfo{author}{\bibfnamefont{A.}~\bibnamefont{Alsolami}}, \bibnamefont{and}
  \bibinfo{author}{\bibfnamefont{J.}~\bibnamefont{Burridge}},
  \bibinfo{journal}{J. Stat. Mech.} p. \bibinfo{pages}{083207}
  (\bibinfo{year}{2018}).

\bibitem[{\citenamefont{Ben-Isaac et~al.}(2015)\citenamefont{Ben-Isaac, Fodor,
  van Wijland, and Gov}}]{BenIsaac2015}
\bibinfo{author}{\bibfnamefont{E.}~\bibnamefont{Ben-Isaac}},
  \bibinfo{author}{\bibfnamefont{E.}~\bibnamefont{Fodor}},
  \bibinfo{author}{\bibfnamefont{F.}~\bibnamefont{van Wijland}},
  \bibnamefont{and} \bibinfo{author}{\bibfnamefont{N.}~\bibnamefont{Gov}},
  \bibinfo{journal}{Phys. Rev. E} \textbf{\bibinfo{volume}{92}},
  \bibinfo{pages}{12716} (\bibinfo{year}{2015}).

\bibitem[{\citenamefont{Razin et~al.}(2019)\citenamefont{Razin, Voituriez, and
  Gov}}]{Razin2019}
\bibinfo{author}{\bibfnamefont{N.}~\bibnamefont{Razin}},
  \bibinfo{author}{\bibfnamefont{R.}~\bibnamefont{Voituriez}},
  \bibnamefont{and} \bibinfo{author}{\bibfnamefont{N.}~\bibnamefont{Gov}},
  \bibinfo{journal}{Phys. Rev. E} \textbf{\bibinfo{volume}{99}},
  \bibinfo{pages}{022419} (\bibinfo{year}{2019}).

\bibitem[{\citenamefont{Magdziarz}(2017)}]{Magdziarz2017}
\bibinfo{author}{\bibfnamefont{M.}~\bibnamefont{Magdziarz}},
  \bibinfo{journal}{Phys. Rev. E} \textbf{\bibinfo{volume}{95}},
  \bibinfo{pages}{022126} (\bibinfo{year}{2017}).

\bibitem[{\citenamefont{Ben-Isaac et~al.}(2011)\citenamefont{Ben-Isaac, Park,
  Popescu, Brown, Gov, and Shokef}}]{BenIsaac2011}
\bibinfo{author}{\bibfnamefont{E.}~\bibnamefont{Ben-Isaac}},
  \bibinfo{author}{\bibfnamefont{Y.}~\bibnamefont{Park}},
  \bibinfo{author}{\bibfnamefont{G.}~\bibnamefont{Popescu}},
  \bibinfo{author}{\bibfnamefont{F.}~\bibnamefont{Brown}},
  \bibinfo{author}{\bibfnamefont{N.}~\bibnamefont{Gov}}, \bibnamefont{and}
  \bibinfo{author}{\bibfnamefont{Y.}~\bibnamefont{Shokef}},
  \bibinfo{journal}{Phys. Rev. Lett.} \textbf{\bibinfo{volume}{106}},
  \bibinfo{pages}{238103} (\bibinfo{year}{2011}).

\bibitem[{\citenamefont{Weisstein}({\natexlab{a}})}]{hypergeometric}
\bibinfo{author}{\bibfnamefont{E.~W.} \bibnamefont{Weisstein}},
  \emph{\bibinfo{title}{{Confluent Hypergeometric Function of the First Kind.
  From MathWorld---A Wolfram Web Resource}}},
  \urlprefix\url{http://mathworld.wolfram.com/ConfluentHypergeometricFunctionoftheFirstKind.html}.

\bibitem[{\citenamefont{Reynolds et~al.}(2013)\citenamefont{Reynolds, Lepretre,
  and Bohan}}]{Reynolds2013}
\bibinfo{author}{\bibfnamefont{A.}~\bibnamefont{Reynolds}},
  \bibinfo{author}{\bibfnamefont{L.}~\bibnamefont{Lepretre}}, \bibnamefont{and}
  \bibinfo{author}{\bibfnamefont{D.}~\bibnamefont{Bohan}},
  \bibinfo{journal}{Sci. Rep.} \textbf{\bibinfo{volume}{3}},
  \bibinfo{pages}{3158} (\bibinfo{year}{2013}).

\bibitem[{\citenamefont{Palyulin et~al.}(2017)\citenamefont{Palyulin,
  Mantsevich, Klages, Metzler, and Chechkin}}]{Palyulin2017}
\bibinfo{author}{\bibfnamefont{V.}~\bibnamefont{Palyulin}},
  \bibinfo{author}{\bibfnamefont{R.}~\bibnamefont{Mantsevich}},
  \bibinfo{author}{\bibfnamefont{R.}~\bibnamefont{Klages}},
  \bibinfo{author}{\bibfnamefont{R.}~\bibnamefont{Metzler}}, \bibnamefont{and}
  \bibinfo{author}{\bibfnamefont{V.}~\bibnamefont{Chechkin}},
  \bibinfo{journal}{Eur. Phys. J. B} \textbf{\bibinfo{volume}{90}},
  \bibinfo{pages}{170} (\bibinfo{year}{2017}).

\bibitem[{\citenamefont{Johnson et~al.}(2008)\citenamefont{Johnson, London,
  Lea, and Durban}}]{Johnson2008}
\bibinfo{author}{\bibfnamefont{D.}~\bibnamefont{Johnson}},
  \bibinfo{author}{\bibfnamefont{J.}~\bibnamefont{London}},
  \bibinfo{author}{\bibfnamefont{M.}~\bibnamefont{Lea}}, \bibnamefont{and}
  \bibinfo{author}{\bibfnamefont{J.}~\bibnamefont{Durban}},
  \bibinfo{journal}{Ecology} \textbf{\bibinfo{volume}{89}},
  \bibinfo{pages}{1208} (\bibinfo{year}{2008}).

\bibitem[{\citenamefont{Kurihara et~al.}(2017)\citenamefont{Kurihara, Aridome,
  Ayade, Zaid, and Mizuno}}]{Kurihara2017}
\bibinfo{author}{\bibfnamefont{T.}~\bibnamefont{Kurihara}},
  \bibinfo{author}{\bibfnamefont{M.}~\bibnamefont{Aridome}},
  \bibinfo{author}{\bibfnamefont{H.}~\bibnamefont{Ayade}},
  \bibinfo{author}{\bibfnamefont{I.}~\bibnamefont{Zaid}}, \bibnamefont{and}
  \bibinfo{author}{\bibfnamefont{D.}~\bibnamefont{Mizuno}},
  \bibinfo{journal}{Phys. Rev. E} \textbf{\bibinfo{volume}{95}},
  \bibinfo{pages}{030601} (\bibinfo{year}{2017}).

\bibitem[{\citenamefont{Mwaffo et~al.}(2015)\citenamefont{Mwaffo, Anderson,
  Butail, and Porfirii}}]{Mwaffo2015}
\bibinfo{author}{\bibfnamefont{V.}~\bibnamefont{Mwaffo}},
  \bibinfo{author}{\bibfnamefont{R.}~\bibnamefont{Anderson}},
  \bibinfo{author}{\bibfnamefont{S.}~\bibnamefont{Butail}}, \bibnamefont{and}
  \bibinfo{author}{\bibfnamefont{M.}~\bibnamefont{Porfirii}},
  \bibinfo{journal}{J. R. Soc. Interface} \textbf{\bibinfo{volume}{12}},
  \bibinfo{pages}{20140884} (\bibinfo{year}{2015}).

\bibitem[{\citenamefont{Weisstein}({\natexlab{b}})}]{hermite}
\bibinfo{author}{\bibfnamefont{E.~W.} \bibnamefont{Weisstein}},
  \emph{\bibinfo{title}{{Hermite Polynomial. From MathWorld---A Wolfram Web
  Resource}}},
  \urlprefix\url{http://mathworld.wolfram.com/HermitePolynomial.html}.

\bibitem[{\citenamefont{Weisstein}({\natexlab{c}})}]{erf}
\bibinfo{author}{\bibfnamefont{E.~W.} \bibnamefont{Weisstein}},
  \emph{\bibinfo{title}{{Erf. From MathWorld---A Wolfram Web Resource}}},
  \urlprefix\url{http://mathworld.wolfram.com/Erf.html}.

\end{thebibliography}

\end{document}